\documentclass[a4paper,11pt]{article}
\pdfoutput=1 

\usepackage{jcappub} 

\newcommand{\T}{\mathbf{\hat{\mathcal{T}}}}

\newcommand{\mI}{\mathcal{I}}

\newcommand{\Dk}[1]{\frac{d^3#1}{(2\pi)^3}}

\newcommand{\ve}[1]{{\text{\bf #1}}} 
\newcommand{\vk}{\ve k}
\newcommand{\vp}{\ve p}
\newcommand{\vq}{\ve q}
\newcommand{\vx}{\ve x}

\newcommand{\A}{\mathcal{A}}
\newcommand{\B}{\mathcal{B}}
\newcommand{\mA}{\mathcal{A}}
\newcommand{\mB}{\mathcal{B}}

\newcommand{\obd}{\omega_\text{BD}}

\newcommand{\ikk}{\underset{\vk_{12}= \vk}{\int}}

\newcommand{\dD}{\delta_\text{D}}
\newcommand{\Ps}{\mathbf{\Psi}}

\title{\boldmath Nonlinear evolution of initially biased tracers in modified gravity}

\author[a,b]{Alejandro Aviles}
\emailAdd{avilescervantes@gmail.com}

\author[b]{Mario Alberto Rodriguez-Meza}
\emailAdd{marioalberto.rodriguez@inin.gob.mx}

\author[a,b,c]{Josue De-Santiago}
\emailAdd{jsantiago@fis.cinvestav.mx}

\author[b]{Jorge L. Cervantes-Cota}
\emailAdd{jorge.cervantes@inin.gob.mx}

\affiliation[a]{Consejo Nacional de Ciencia y Tecnolog\'ia, Av. Insurgentes Sur 1582,
Colonia Cr\'edito Constructor, Del. Benito Jurez, 03940, Ciudad de M\'exico, M\'exico}
\affiliation[b]{Departamento de F\'isica, Instituto Nacional de Investigaciones Nucleares,
Apartado Postal 18-1027, Col. Escand\'on, Ciudad de M\'exico,11801, M\'exico.}
\affiliation[c]{Depto. de F\'isica, Centro de Investigaci\'on y de Estudios Avanzados del IPN, A.P. 14-740, 
07000 Ciudad de M\'exico, M\'exico}


\abstract{
In this work we extend the perturbation theory for modified gravity (MG) in two main aspects. 
First, the construction of matter overdensities
from Lagrangian displacement fields is shown to hold in a general framework, allowing us 
to find Standard Perturbation Theory (SPT) kernels from known Lagrangian Perturbation Theory (LPT) kernels. 
We then develop a theory of biased tracers for generalized cosmologies, extending already existing formalisms 
for $\Lambda$CDM.
We present the correlation function in Convolution-LPT and the power spectrum in SPT for $\Lambda$CDM, $f(R)$ Hu-Sawicky, 
and DGP braneworld models.
Our formalism can be applied to many generalized cosmologies and to facilitate it, we are making public a code to compute these statistics.
We further study the relaxation of bias with the use of a simple model and of excursion set theory, showing that in general the bias
parameters have smaller values in MG than in General Relativity.
}

\begin{document}
\maketitle
\flushbottom

\begin{section}{Introduction}

With the advent of large scale structure surveys such as Euclid\footnote{\href{https://www.euclid-ec.org}{www.euclid-ec.org}} \cite{Amendola:2012ys}, 
DESI\footnote{\href{https://www.desi.lbl.gov}{www.desi.lbl.gov}} \cite{Aghamousa:2016zmz}, and 
LSST\footnote{\href{https://www.lsst.org/}{www.lsst.org}} \cite{Abate:2012za}
we will determine the Baryon Acoustic Oscillations (BAO) and Reshift Space Distortions (RSD) in much more precise way than in previous surveys,
and we will be in position to test gravity at unprecedented cosmological scales. 
These upcoming experiments will cover wider an deeper regions of
space, probing more modes that behave linearly and quasi-linearly than in previous surveys. 
Henceforth, the methods of Perturbation Theory (PT) for dark matter clustering will be very valuable for the understanding of the outcomes
of such experiments. As a matter of fact, with the exception of weak lensing observations that directly probes the dark matter distribution, the objects of 
observations will be biased tracers of the underlying matter content. 
Hence, a PT that incorporates biased tracers is needed. The main difficulty of such a theory relies 
on that the formation and evolution of tracers are highly nonlinear processes that cannot be captured by
PT itself. Hence, their effects are commonly accommodated within an effective field theory that integrates out 
small scales, by smoothing the relevant dynamical fields. 
As a result, a set of bias parameters that encode the
ignorance of our theory are incorporated, and should be estimated from observations or simulations; 
or otherwise, they may be obtained from a bias model, as in the peak-background split procedure \cite{Kaiser:1984sw,Mo:1996cn,Sheth:1999mn} applied to 
a universal mass function \cite{Press:1973iz,Sheth:1999mn}. 
An inconvenience of this effective description is that the smoothing scale is arbitrary and the final results 
depend on it, thus a renormalization procedure for the biases should be accounted for
\cite{McDonald:2006mx,McDonald:2009dh,Schmidt:2012ys,Assassi:2014fva,Aviles:2018thp}.  The modeling of bias in $\Lambda$CDM has 
been studied for decades  \cite{Kaiser:1984sw,Fry:1992vr,Fry:1996fg,Matarrese:1997sk,Mo:1996cn,Sheth:1999mn,Matsubara:2008wx,McDonald:2009dh,Matsubara:2011ck} 
and by now is in a mature state \cite{Desjacques:2016bnm}.
For generalized cosmologies the situation is rather different, and comparatively 
little work on the subject has been developed so far \cite{Hui:2007zh,Parfrey:2010uy,Lombriser:2013eza,Munshi:2016zzr}. 
By generalized cosmologies we mean models that contain dark matter and the standard matter
particles, but the 
acceleration of the Universe is driven by other unknown fields, rather than a cosmological constant. 
One may think of the case of dark energy, but this is almost trivial for PT since their effects are
mainly a consequence of the overall Hubble flow, and its impact on the nonlinear regime is 
small because its perturbations grow almost negligible and are stress-free. The case of Modified Gravity (MG) 
is in this sense more interesting since one can easily have a theory observationally indistinguishable 
from $\Lambda$CDM at the background level, but predicting very different clustering 
of matter \cite{Bertschinger:2006aw,Bertschinger:2008zb}. Therefore, in this work we concentrate 
on MG theories; for recent reviews on infrared modifications to General Relativity (GR) see 
\cite{Clifton:2011jh,Joyce:2014kja,Koyama:2015vza}.

Lagrangian Perturbation Theory (LPT) \cite{Zel70,Buc89,Mou91,Catelan:1994ze,BouColHivJus95,Taylor:1996ne,Matsubara:2007wj,Carlson:2012bu,Matsubara:2015ipa} 
is highly successful in modeling the matter correlation function and the phases of the BAO, 
but it poorly models the broadband of the nonlinear power spectrum. 
On the other hand, (Eulerian) Standard Perturbation Theory (SPT) performs better in the broadband power spectrum, but
formally it cannot be Fourier transformed to obtain the correlation function \cite{Bernardeau:2001qr}: if one insists and performs a cutoff to the power
spectrum at some arbitrary, small scale, and Fourier transforms it, the correlation function will show a double peak structure around the
BAO scale \cite{Baldauf:2015xfa}, reflecting the artificial phase shift induced by the mode coupling in the Eulerian formalism. 
In that sense, LPT and SPT approaches are complementary \cite{Tassev:2013rta}. 
The advantage of starting with displacement fields is that matter $2$-point functions can be obtained 
from 2- and 3-point cumulants of the Lagrangian displacements, for both SPT and LPT statistics:
for $\Lambda$CDM model, the LPT power spectrum was shown to be a resummation of that of SPT \cite{Matsubara:2007wj,Vlah:2014nta}. 
In this work we show that this procedure works in more generality, 
by finding equations that relate the SPT and LPT kernels and
that are valid for MG theories, extending the cases constructed in \cite{Matsubara:2011ck,Rampf:2012xa}. This fact may be
not surprising since the relation between Lagrangian and Eulerian frames has a geometric nature, but a formal proof was lacking. 
By doing this, we are able to construct the SPT kernels from known LPT kernels, 
and we present explicit SPT second order kernels in MG, which include coefficients that are
solutions to simple linear second order differential equations, such that they may be solved also by Green methods. 

The nonlinear PT for MG was developed initially in \cite{2009PhRvD..79l3512K} 
based on the closure theory for structure formation of ref.~\cite{Taruya:2007xy}, and further studied in several other works 
\cite{Taruya:2013quf,Brax:2013fna,Taruya:2014faa,Bellini:2015oua,Taruya:2016jdt,Bose:2016qun,Fasiello:2017bot,Bose:2017dtl,Aviles:2017aor,Hirano:2018uar,Bose:2018orj,Bose:2018zpk}. 
But bias has been considered only at the linear level \cite{Hui:2007zh,Parfrey:2010uy,Lam:2012fa,Lombriser:2013eza}, 
with the exception of time dependent parametrized models \cite{Munshi:2016zzr}.
In this work we are aiming to fill this gap by considering the 
nonlinear evolution of initially biased tracers. We will focus on the structure of the PT, 
instead than on the evolution of the bias parameters themselves. But for the latter we put forward a simple model 
that, although primitive, captures the fact that for theories that incorporate additional scalar fields to the gravity sector, 
as in $f(R)$ gravity, the local bias parameters acquire smaller values than in GR. 
This effect has been observed recently in simulations \cite{Arnold:2018nmv}, and it is expected since the associated, attractive
fifth force leads to a more efficient clustering of matter, and consequently it provokes a more rapid relaxation of bias.

The LPT for dark matter fluctuations in MG was developed 
in \cite{Aviles:2017aor}, and in the present work we extend that formalism 
to incorporate biased tracers using the tools of ref.~\cite{Aviles:2018thp}. Alternatively, this work can be
considered as an extension of the works of Matsubara \cite{Matsubara:2007wj}, and of Carlson, Reid and White \cite{Carlson:2012bu},  
that introduce the effects of MG; however, here we do not deal with RSD.

Our approach is that of Lagrangian bias, in which one considers an almost 
homogeneous distribution of matter at an early initial time that is smoothed over some scale $R_\Lambda$, and to which the bias procedure
is applied; after that,
the nonlinear evolution of fields takes place. An alternative route is that of Eulerian bias, in which 
the bias is introduced into the evolved fields. We notice  both methods are not equivalent since the process of smoothing
and nonlinear evolution do not commute; thus, for example, initially local bias evolves into non local bias \cite{Matsubara:2011ck}. 
We consider bias operators constructed as powers of the matter overdensity $\delta$ and its curvature
$\nabla^2 \delta$, as in \cite{Schmidt:2012ys,Aviles:2018thp}. 
Our definition for bias parameters will be that proposed in ref.~\cite{Aviles:2018thp}, as an extension to the one
introduced in ref.~\cite{Matsubara:2008wx}. Once renormalized, these bias parameters
are equivalent to the peak-background split biases, which quantify the change in the 
tracer abundances against variations of the background density and its curvature. 
Other possible operators as tidal bias will not be treated here since they are generated by the nonlinear 
evolution \cite{Baldauf:2012hs}, even if they are not initially present. 
Nevertheless, it is worth to notice that there are some important evidences of the existence 
of a non-zero Lagrangian tidal bias \cite{Modi:2016dah,Vlah:2016bcl}. 
Our formalism can be extended to include Lagrangian tidal bias, as in ref.\cite{Vlah:2016bcl}, 
but the renormalization procedure, under our bias prescription, is not clear. 
Also, in our formalism we introduce operators constructed from the new fields present in the MG theories, specifically
the Laplacian of the scalar field; but we will observe that this is degenerated with 
the curvature bias for scales larger than the range of the induced fifth force.
Therefore, we keep the prescription to use only a Lagrangian bias function $F$ 
with two arguments that relates matter and tracers ($X$) overdensities as 
$1+\delta_X = F(\delta,\nabla^2 \delta)$. 
  A more rigorous approach would construct all the invariant operators 
 out of the fields entering the theory 
 up to the desired order in fluctuations, and thereafter perform the bias expansion. However, we note this is no sufficient for general MG models 
 because the linear growth cannot be factorized in scale and time dependent pieces and hence such expansion is not complete (see for example the discussion
 in sect.~8.3 of \cite{Desjacques:2016bnm}). However, by noting that the source of the Klein-Gordon equation can be expanded in powers of $k^2/m^2 a^2$ this 
 drawback in our formalism is partially tamed by operators $\nabla^2 \delta$, $\nabla^4 \delta$, and so on, for sufficiently large scales.

We present results for the correlation function in the context of Convolution Lagrangian Perturbation Theory (CLPT) \cite{Carlson:2012bu} 
and the power spectrum in SPT.  
Throughout this work we exemplify our findings with $\Lambda$CDM,  Hu-Sawicky $f(R)$ \cite{Hu:2007nk,Starobinsky:2007hu} 
and the Dvali-Gabadadze-Porrati (DGP) braneworld \cite{Dvali:2000hr} models;
although, as in several previous works, the techniques developed here 
can be applied to other MG models, essentially to the whole Horndeski sector \cite{Horndeski:1974wa}, as shown in ref.~\cite{Bose:2016qun}.

The rest of the paper is organized as follows: In section \ref{sec:2} we first review the LPT formalism in MG and 
thereafter we find relations among the SPT and LPT kernels; in section \ref{sec:3} we present a formalism 
for initially biased tracers and their nonlinear evolution, obtaining the structure of the matter 2-point functions; 
in section \ref{sec:4} we put forward a simplified model for the estimation of the numerical values of the bias parameters, and compare 
them to the outputs of excursion set theory with moving barriers; finally, we conclude
in section \ref{sec:conclusions}. Almost all lengthy computations are delegated to appendices; 
specifically, in appendices \ref{app:kqfunctions} and \ref{app:PS} we construct all the necessary 
functions to compute the power spectra and correlation functions.

\end{section}

\begin{section}{Nonlinear evolution of matter fluctuations}\label{sec:2}

\begin{subsection}{Lagrangian displacement fields in generalized cosmologies}\label{sec:2.1}

This subsection reviews the LPT for MG formalism, and also helps us to set our notation; for details see \cite{Aviles:2017aor}.

Matter particles follow trajectories with Eulerian comoving coordinates $\vx$. The Lagrangian displacement vector field $\Ps$ 
relates the initial (Lagrangian) $\vq$ and Eulerian $\vx$ positions of particles as
\begin{equation}\label{SPTtoLPTcoord}
\Ps(\vq,t) =  \vx(\vq,t) - \vq + \mathbf{\Gamma}(\vq,t),
\end{equation}
chosen such that $\vx(\vq,t_{ini})= \vq$, with $t_{ini}$ an early time where the evolution of all scales of interest remains linear and 
the overdensities are quite small, $\delta(\vx, t_{ini}) = \delta(\vq) \ll 1$. We introduced a vector $\mathbf{\Gamma}$ that carries the 
transverse piece of the Lagrangian displacement, for irrotational perfect fluids it has contributions 
starting at third order in perturbation theory \cite{Matsubara:2015ipa}, but they do not show up in matter density 2-point, 1-loop statistics. 
The vector $\mathbf{\Gamma}$, on the other hand, has contributions at all orders if we aim to describe the dispersion of the velocity of 
particles \cite{Aviles:2015osc} and the generation of vorticity \cite{Cusin:2016zvu}. 
In this work the matter content is composed only by cold dark matter particles and we want to describe the lowest corrections to statistics, 
allowing us to safely set $\mathbf{\Gamma}=0$, and to consider $\Ps$ a longitudinal field. Using matter conservation, 
\begin{equation}\label{mattercons}
(1 + \delta(\vx,t))d^3 x = (1 + \delta(\vq))d^3 q, 
\end{equation}
one gets the known relation between density and Lagrangian fields 
\begin{equation}
 \delta(\vx,t) = \frac{1 + \delta(\vq)-J}{J} \simeq  \frac{1 - J }{J},
\end{equation}
where $J_{ij} = \partial x_i/\partial q^j = \delta_{ij} + \Psi_{i,j}$ is the Jacobian matrix of 
the coordinate transformation in eq.~(\ref{SPTtoLPTcoord}) and $J$ its determinant. Since $\Ps$ is a potential field, 
the Jacobian matrix is symmetric.

A main generic feature found in models that modify gravity is that the effective gravitational strength ``$G_{eff}$'' 
becomes scale and time dependent. In wide generality, the linearized fluid equations take the form 
\begin{equation}\label{lineardeltaeq}
 (\T -  A(k,t)) \delta_L(k,t) = 0,   
\end{equation}
where $\T \equiv \frac{d^2\,}{dt^2} + 2 H \frac{d\,}{dt}$, as was introduced in \cite{Matsubara:2015ipa}, and we defined
\begin{align}
 A(k,t) &= A_0  \left( 1 + \frac{2 \beta^2 k^2}{k^2 + m^2 a^2} \right), \label{Afunct}\\ 
  A_0 &= 4 \pi G \bar{\rho}, \label{A0}
\end{align}
with $\bar{\rho}$ the background matter density.
In general $\beta$ and $m$ are time and scale dependent and may be viewed as arbitrary parameters for unknown underlying 
theories \cite{Bertschinger:2008zb,Silvestri:2013ne}.
Or, otherwise, they can be obtained directly from a specific gravitational model. 
As long as $m\neq 0$, for scales $k/a \ll m$ we recover GR, while for small scales 
we get  $A = (1+2 \beta^2) A_0 $. For example, in $f(R)$ gravity $\beta = -1/\sqrt{6}$, meaning that
the strength of the gravitational interaction is enhanced by a factor $4/3$ at small scales. 
Clearly this fact would invalidate any $f(R)$ model if it were not for the 
presence of the nonlinearities of the theory, below encoded in the term $\delta I$, 
leading to the chameleon effect and driving the theory
to GR in the appropriate limits.
A notable example in which the function $A$ is scale independent is the DGP model, 
for which the mass is zero, though linear growth is affected by a time dependent $\beta^2$ function. 
Our formalism deals with this model by simply setting 
$m=0$ in eq.~(\ref{Afunct}). But note that it does not reduce to GR at large scales, implying that DGP is
highly constrained by background cosmology observations. In the following we will keep the discussion general 
and assume that the function $A$ is $k$-dependent. Because of this, the solutions to
eq.~(\ref{lineardeltaeq}), named the linear growth 
functions,  carry a $k$ dependence as well. We will denote the 
fastest growing of these two solutions as  $D_+(k,t)$, and we omit the discussion of the
decaying solution.\footnote{Unlike \cite{Aviles:2017aor} that uses Green function methods, here we solve for the 
higher order growth functions with differential equations, which we find numerically simpler. If
the former method is used, it is necessary to consider both growing and decaying solutions to eq.~(\ref{lineardeltaeq}).}

The equation of motion for the Lagrangian displacement is given by the geodesic equation,
\begin{equation} \label{EoMLD}
 \T \Ps(\vq,t) = - \frac{1}{a}\nabla_\vx \psi(\vx,t) |_{\vx=\vq + \Ps}. 
\end{equation}
We use $\nabla_\vx = \partial/\partial \vx$ to denote 
partial differentiation with respect to Eulerian coordinates, and  $\nabla = \partial/\partial \vq$ for differetiantion with respect to
Lagrangian coordinates.
The two gravitational potentials of the metric 
are related by $\phi -\psi = \varphi/2$, where $\varphi$ is the scalar field mediating the fifth force, such that
the Poisson equation is modified to
\begin{equation}\label{poissoneq}
\frac{1}{a^2}\nabla^2_\vx \psi = A_0 \delta(\vx,t) - \frac{1}{2 a^2}\nabla^2_\vx \varphi. 
\end{equation}
The scalar field is coupled to the trace of the matter energy momentum 
tensor, $T = -\rho$, and its Klein-Gordon equation in the quasi-static limit 
takes the form \cite{2009PhRvD..79l3512K}
\begin{equation}\label{KGeq}
 \frac{1}{a^2}\nabla^2_\vx \varphi = -  4 A_0 \beta^2 \delta + m^2 \varphi + 2 \beta^2 \delta \mathcal{I},
\end{equation}
where $\delta I$ is defined to contain all nonlinear terms,  and here we expand it in Fourier space
as
\begin{align} \label{deltaI}
 \delta \mathcal{I}(\vk) &= \frac{1}{2} \int \frac{d^3 k_1d^3 k_2}{(2\pi)^3} \dD (\vk-\vk_1-\vk_2) M_2(\vk_1,\vk_2) \varphi(\vk_1)\varphi(\vk_2) \nonumber\\
 &+ \frac{1}{3}  \int \frac{d^3 k_1d^3 k_2 d^3k_3}{(2\pi)^6} \dD (\vk-\vk_1-\vk_2-\vk_3) M_3(\vk_1,\vk_2,\vk_3) \varphi(\vk_1)\varphi(\vk_2)\varphi(\vk_3).
\end{align}
Equivalent expansions on matter overdensities are also common in the literature. In Brans-Dicke 
theories we identify $3+2 \obd =1/ 2 \beta^2$, allowing us to recognize the $\beta$ parameter as the strength 
of the matter to scalar field coupling. Instead of the mass $m$, other works 
use the notation $M_1  = m^2/2 \beta^2 $, making eq.~(\ref{KGeq}) more symmetrical. The combination 
$\mathcal{I} = M_1 \varphi + \delta \mathcal{I}$ corresponds to the derivative of the scalar field potential.

It is convenient to recast  eq.~(\ref{EoMLD}) in terms of only Lagrangian
coordinates, then by taking the $\vx$-divergence and transforming to Lagrangian coordinates with 
the aid of $\nabla_{\vx i} = (J^{-1})_{ji} \nabla_{i}$, we get in Fourier space
\begin{equation} \label{preEqM}
 \big[ (J^{-1})_{ij} \T \Psi_{i,j} \big](\vk) = - A(k) \tilde{\delta}(\vk) 
 + \frac{2 \beta^2 k^2}{k^2+m^2 a^2} \delta I (\vk)  +   \frac{1}{2} \frac{m^2}{k^2+m^2 a^2}   [(\nabla^2_\vx \varphi - \nabla^2 \varphi)](\vk).
\end{equation}
We have used the notation $[(\cdots)](\vk)$ to denote the Fourier transform of $(\cdots)(\vq)$, and 
\begin{equation}
\tilde{\delta}(\vk) = \int d^3 q e^{-i\vq \cdot \vk} \delta(\vx) = \left[\frac{1-J(\vq)}{J(\vq)}\right](\vk). 
\end{equation}
The geometric term $[(\nabla^2_\vx \varphi - \nabla^2 \varphi)](\vk)$ is called frame-lagging and is more
important at large scales, especially if the theory is 
expected to reduce to GR in that limit. It arises from the correction of spatial derivatives when 
transforming from the Eulerian to the Lagrangian frame. 
Equation (\ref{preEqM}) is solved perturbatively using  
$(J^{-1})_{ij} = \delta_{ij} - \Psi_{i,j} + \Psi_{i,k}\Psi_{k,j} + \cdots$. To linear order we have
\begin{equation}
  (\T -  A(k,t)) [\Psi^{(1)}_{i,i}] = 0, 
\end{equation}
which is the same equation for the linear matter overdensity. Hence, the first order solution is
\begin{equation} \label{Psi1}
 \Psi^{(1)}_i(\vk,t) = i \frac{k_i}{k^2} \delta_L(\vk,t) =  i \frac{k_i}{k^2} D_+(k,t) \delta_L(\vk,t_0),
\end{equation}
where we notice the normalization is fixed by eq.~(\ref{mattercons}), 
and  we choose the growth function such that $D_+(k \rightarrow 0,t_0)=1$, where $t_0$ denotes present time.
More generally, Lagrangian displacements are expanded as
\begin{equation} \label{PsiExp}
 \Psi_i(\vk) = \sum_{n=0}^{\infty} \Psi^{(n)} =  \sum_{n=1}^{\infty} \frac{i}{n!}\underset{\vk_{1 \dots n}= \vk}{\int} 
 L_i^{(n)}(\vk_1,...,\vk_n) \delta_L(\vk_1) \cdots \delta_L(\vk_n). 
\end{equation}
Hereafter we make use of the shorthand notation
\begin{equation}
 \underset{\vk_{1\cdots n}= \vk}{\int} = \int \Dk{\vk_1} \cdots \Dk{\vk_n} (2 \pi)^3 \delta_\text{D}(\vk - \vk_{1\cdots n}),
\end{equation}
and  $\vk_{1\cdots n} = \vk_1 + \cdots +\vk_n$ denotes the sum of arbitrary number of momenta.

The kernels $L^{(n)}_i$ are obtained order by order using eq.~(\ref{preEqM}). 
At first order, reading eq.~(\ref{Psi1}) above, we have 
\begin{equation}\label{LPTkern1}
L_i^{(1)}(\vk) = \frac{k_i}{k^2}, 
\end{equation}
while to second order we obtain \cite{Aviles:2017aor}
\begin{equation}\label{LPTkern2}
L_i^{(2)}(\vk_1,\vk_2) = \frac{3}{7} \frac{k_i}{k^2}\left( \mA(\vk_1,\vk_2) - \mB(\vk_1,\vk_2) \frac{(\vk_1\cdot \vk_2)^2}{k_1^2 k_2^2} \right) , 
\end{equation}
with $\vk = \vk_1 + \vk_2$,
\begin{equation} \label{AandBdef}
 \mA(\vk_1,\vk_2) = \frac{7 D^{(2)}_{\mA}(\vk_1,\vk_2)}{3 D_{+}(k_1)D_{+}(k_2)}, 
 \qquad \mB(\vk_1,\vk_2) = \frac{7 D^{(2)}_{\mB}(\vk_1,\vk_2)}{3 D_{+}(k_1)D_{+}(k_2)},
\end{equation}
and the second order growth functions are the solutions to equations
\begin{align}
(\T - A(k))D^{(2)}_{\mA} &= \Bigg[A(k) + (A(k)-A(k_1))\frac{\vk_1\cdot\vk_2}{k_2^2} + (A(k)-A(k_2))\frac{\vk_1\cdot\vk_2}{k_1^2} \nonumber\\
             &     \qquad   -  \left(\frac{2 A_0}{3}\right) \frac{k^2}{a^2} \frac{M_2(\vk_1\vk_2)}{6 \Pi(k)\Pi(k_1)\Pi(k_2)}\Bigg]  D_{+}(k_1)D_{+}(k_2), \label{DAeveq} \\
(\T - A(k))D^{(2)}_{\mB} &= \Big[A(k_1) + A(k_2) - A(k) \Big]  D_{+}(k_1)D_{+}(k_2), \label{DBeveq}
\end{align}
with appropriate initial conditions. As it is common, we have used $\Pi(k) \equiv (k^2 + m^2a^2)/6\beta^2a^2$. 
The second and third terms in the right hand side of eq.~(\ref{DAeveq}) arise because of the frame-lagging contributions. 
The fourth term is the second order contribution of $\delta I$, which in MG is responsible of the screening mechanism. 
Since $\mA$ and $\mB$  depend on $\vk_1$ and $\vk_2$, the decomposition in eq.~(\ref{LPTkern2}) is arbitrary and we adopt it because 
they take values of order unity and the connection to $\Lambda$CDM is direct. For this case we obtain 
\begin{align}
D^{(2)}_{\mA}(t) = D^{(2)}_{\mB}(t) &=  (\T-A_0)^{-1}\left[ \frac{3}{2}\Omega_m H^2 D_+^2 \right] \nonumber\\
&= \frac{3}{7}D_+^2(t) + \frac{4}{7} (\T-A_0)^{-1}\left[ \frac{3}{2}\Omega_m H^2 D_+^2 \left(1-\frac{f^2}{\Omega_m} \right)\right],
\end{align}
thus $\A=\B$ are only time dependent. For $f=\Omega_m^{1/2}$ we get $\A^\text{EdS}=\B^\text{EdS}=1$ 
and the standard kernels in Einstein-de Sitter (EdS) are recovered.

The third order kernel $L^{(3)}_i(\vk_1,\vk_2,\vk_3)$ is provided in ref.~\cite{Aviles:2017aor} and we do not reproduce it here.

 
\end{subsection}

\begin{subsection}{From Lagrangian to Standard Perturbation Theory}\label{sec:2.2}

An alternative approach is to deal from the beginning with the overdensity $\delta(\vx)$ 
and the velocity divergence $\theta(\vx) = \nabla_\vx \cdot \ve v / (a H)$
fields that evolve according to the hydrodynamical continuity and Euler equations, in Fourier space  
 \begin{align} 
 H^{-1}\partial_t \delta(\vk) + \theta(\vk) &= -\ikk \alpha(\vk_1,\vk_2)\theta(\vk_1) \delta(\vk_2), \label{ContEq}\\
 H^{-1}\partial_t \theta(\vk) +  \left( 2+\frac{\dot{H}}{H^2}\right)\theta(\vk) - \left( \frac{k}{aH}\right)^2\psi(\vk) &= 
  -\frac{1}{2} \ikk \beta(\vk_1,\vk_2)\theta(\vk_1) \theta(\vk_2) \label{EulerEq},
\end{align}
respectively, supplemented by the Poisson equation (\ref{poissoneq}). The functions $\alpha$ and $\beta$ describe how
two plane waves interact and are given by
\begin{equation}
\alpha(\vk_1,\vk_2) = 1 +\frac{\vk_1 \cdot \vk_2}{2}\left(\frac{1}{k_1^2} + \frac{1}{k_2^2} \right), \qquad \beta(\vk_1,\vk_2) = \frac{(\vk_1 \cdot \vk_2)|\vk_1 + \vk_2 |^2}{k_1^2 k_2^2}.
\end{equation}
In SPT the expansion is performed directly to the overdensity and velocity fields, $\delta = \delta^{(1)} + \delta^{(2)}+\cdots$ and
$\theta = \theta^{(1)} + \theta^{(2)}+\cdots$, which in Fourier space can be written as
 \begin{align} 
 \delta^{(n)}(\vk) &= \underset{\vp_1+\cdots+\vp_n=\vk}{\int}F_n(\vp_1,\dots,\vp_n)\delta_L(\vp_{1}) \cdots \delta_L(\vp_{n}), \label{deltan}\\
 \theta^{(n)}(\vk) &= - \underset{\vp_1+\cdots+\vp_n=\vk}{\int}G_n(\vp_1,\dots,\vp_n)\delta_L(\vp_{1}) \cdots \delta_L(\vp_{n}). \label{deltatheta}
 \end{align}
The $F_n$ and $G_n$ kernels are obtained by solving iteratively eqs.~(\ref{ContEq}, \ref{EulerEq}).
In this notation, the linear growth functions $D_+$ are 
kept attached to the linear fields because they are scale dependent and cannot be pulled out of the integral;
and the $G_n$ kernels carry the linear growth factors $f=d \log D_+ / d \log a$. 
Our notation coincides with 
that of ref.~\cite{Taruya:2016jdt} except for a minus sign in $G_n$, but differs from the most used notations that factorize the $f$ factors. 
In the following we obtain these kernels from the Lagrangian kernels at arbitrary perturbative order.

From matter conservation (\ref{mattercons}),
the Lagrangian displacement field is related to the matter density field by
\begin{equation}\label{delta1}
 \delta(\vk,t) = \int d^3 q e^{- i \vk \cdot \vq} \left( e^{-i \vk \cdot \Ps(\vq,t)} -1 \right) = \int d^3 q e^{- i \vk \cdot \vq} 
 \sum_{n=1}^{\infty} \frac{1}{n!} (-i \vk \cdot \Ps(\vq,t))^{n}.
\end{equation}
The inverse $q$-Fourier transform is $\Ps(\vq,t) = \int \Dk{p} e^{i\vp \cdot \vq} \Ps(\vp,t)$, hence
\begin{align} \label{delta2}
 \delta(\vk) = \sum_{\ell=1}^{\infty} \frac{(-i)^\ell}{\ell!} k_{i_1}\cdots k_{i_m} 
 \underset{\vk_{1 \dots \ell}= \vk}{\int} \Psi_{i_1}(\vk_1)\cdots \Psi_{i_\ell}(\vk_\ell),
\end{align}
that is obtained after performing the $\vq$ integration that yields a Dirac delta function that ensures conservation of momentum. 
We now substitute eq.~(\ref{PsiExp}) into each Lagrangian displacement appearing in eq.~(\ref{delta2}) to obtain
\begin{align} \label{delta3}
 \delta(\vk) &= \sum_{\ell=1}^{\infty} \sum_{m_1=1}^{\infty} \cdots \sum_{m_\ell=1}^{\infty} \frac{ k_{i_1}\cdots k_{i_\ell} }{\ell!m_1!\cdots m_\ell!}
 \underset{\vk_{1} + \cdots + \vk_\ell = \vk}{\int} \quad
 \underset{\vp_{11} + \cdots +\vp_{1 m_1} = \vk_1}{\int} \cdots \underset{\vp_{\ell 1}+ \cdots + \vp_{\ell m_\ell}= \vk_\ell}{\int} \nonumber\\
    L_{i_1}^{(m_1)}&(\vp_{11},...,\vp_{1 m_1}) \cdots L_{i_\ell}^{(m_\ell)}(\vp_{\ell 1},...,\vp_{\ell m_\ell})
  \delta_L(\vp_{11}) \cdots \delta_L(\vp_{1 m_1}) \cdots \delta_L(\vp_{\ell 1}) \cdots \delta_L(\vp_{\ell m_\ell}). 
\end{align}
We are looking for an expression with the same form as eq.~(\ref{deltan}). 
At order $n$,  there should be $n$ linear density fields,
then  $m_1 + \cdots + m_\ell = n $, and because $1 \leq m_i \leq n$ we must have at most $\ell =n$. 
Having these considerations in mind and relabeling the momenta $\vp_{11},\dots,\vp_{\ell,m_\ell}$ as
$\vp_1,\dots,\vp_n$, we obtain
\begin{align} \label{delta4}
 \delta^{(n)}(\vk) &= \sum_{\ell=1}^{n} \sum_{m_1+ \cdots + m_\ell=n} \frac{  k_{i_1} \cdots k_{i_\ell} }{\ell!m_1!\cdots m_\ell!}
 \underset{\vp_{1\cdots n} = \vk}{\int} 
    L^{(m_1)}_{i_1}(\vp_{1},...,\vp_{m_1})\cdots L^{(m_\ell)}_{i_\ell}(\vp_{m_{\ell - 1}+1},...,\vp_{ m_\ell}) \nonumber\\
  & \qquad \qquad \times \delta_L(\vp_1) \cdots \delta_L(\vp_{n}). 
\end{align}
This expression gives the SPT $F_n$ kernels in terms of the first $n$ LPT kernels
\begin{align} \label{LPTtoSPTFn}
 F_n(\vk_1,\dots,\vk_n) = \sum_{\ell=1}^{n} \sum_{m_1+ \cdots + m_\ell=n} \frac{ k_{i_1} \cdots k_{i_\ell} }{\ell!m_1!\cdots m_\ell!}
  L^{(m_1)}_{i_1}(\vk_{1},...,\vk_{m_1}) \cdots L^{(m_\ell)}_{i_\ell}(\vk_{m_{\ell - 1}+1},...,\vk_{ m_\ell})
\end{align}
with $\vk = \vk_1 + \cdots + \vk_n$. It may be convenient to symmetrize ($s$) over the arguments in the above kernels; we do that 
for $n=2,3$, yielding
\begin{align}
 F_2(\vk_1,\vk_2) &=  \frac{1}{2} \Big( k_iL^{(2)}_i(\vk_1,\vk_2) + k_i k_j L_i^{(1)}(\vk_1)L_j^{(1)}(\vk_2) \Big)\label{LPTtoF2}\\
 F_3^{s}(\vk_1,\vk_2,\vk_3) &= \frac{1}{3!}\Big( k_i L^{(3)s}_i(\vk_1,\vk_2,\vk_3)  
 + k_ik_j (L^{(2)}_i(\vk_1,\vk_2)L^{(1)}_j(\vk_3) + \text{cyclic})\nonumber\\
 &\qquad + k_ik_jk_kL^{(1)}_i(\vk_1)L^{(1)}_j(\vk_2)L^{(1)}_k(\vk_3) \Big)\label{LPTtoF3}
\end{align}
The two above special cases of eq.~(\ref{LPTtoSPTFn}) were found in  refs.~\cite{Matsubara:2011ck,Rampf:2012xa}.
An explicit expression for $F_2$ is found by substituting eqs.~(\ref{LPTkern1}, \ref{LPTkern2}) into eq.~(\ref{LPTtoF2}):
\begin{align}\label{F2MG}
 F_2(\vk_1,\vk_2) &= \frac{1}{2} + \frac{3}{14}\mA + \left(\frac{1}{2} - \frac{3}{14}\mB \right)\frac{(\vk_1\cdot\vk_2)^2}{\vk_1^2\vk_2^2} 
 + \frac{\vk_1 \cdot \vk_2}{2} \left( \frac{1}{k^2_1} + \frac{1}{k_2^2}\right). 
\end{align}
This equation generalizes eq.~(71) of reference \cite{Bernardeau:2001qr} (after correcting a typo). From here we can identify
the parameter $\epsilon \approx \frac{3}{7}\Omega_m^{2/63}$ \cite{Bouchet:1992uh,Bouchet:1994xp} 
with $\mA^\text{$\Lambda$CDM} =\mB^\text{$\Lambda$CDM} = \frac{7}{3}\epsilon$.

A standard approach to find the SPT kernels, by using Euler and continuity equations, is surprisingly difficult.
However, by knowing the structure of eq.~(\ref{F2MG}), 
we can insert it as ansatz into the hydrodynamical equations and find that $\A$ and $\B$ 
coincide with eqs.~(\ref{AandBdef}). This computation also shows that the frame-lagging terms, that in 
LPT are a consequence of transforming from Eulerian to Lagrangian spatial derivatives, 
in SPT arise as nonlinear responses of velocity fields to the scale- and time-dependent strength. That same approach shows the $G_2$ kernel is
\begin{align}
 G_2(\vk_1,\vk_2) &=  \frac{3}{14}\mA(f_1+f_2) + \frac{3 \dot{\mA}}{14 H} + \left(\frac{f_1+f_2}{2} - \frac{3}{14}\mB(f_1+f_2) 
 - \frac{3 \dot{\mB}}{14 H}\right)\frac{(\vk_1\cdot\vk_2)^2}{\vk_1^2\vk_2^2} \nonumber\\
 &\quad + \frac{\vk_1 \cdot \vk_2}{2} \left( \frac{f_2}{k^2_1} + \frac{f_1}{k_2^2}\right), \label{G2MG}
\end{align} 
where $f_{1,2}=f(k_{1,2})$.
In $\Lambda$CDM, the terms $\dot{\mA}/{H}$ and $\dot{\mB}/{H}$ in $G_2$ can be safely neglected  because 
$\frac{7}{3}\epsilon \simeq 1$ changes in about $1 \%$ over a Hubble time. If we do that, we recover 
eq.~(72) of reference \cite{Bernardeau:2001qr}.
In appendix \ref{app:Gn} we obtain an analogous expression to eq.~(\ref{LPTtoSPTFn}) 
that relates LPT kernels to the SPT $G_n$ kernels; that relation confirms the validity of eq.~(\ref{G2MG}). 

Now, we turn to compute the SPT power spectrum from the LPT formalism. The 1-loop expression is  
\begin{equation}
 P^\text{SPT}_{1\text{-loop}}(k) = P_L(k) + P_{22}(k) + P_{13}(k)
\end{equation}
with $P_{L}(k) = \langle \delta^{(1)}(\vk)\delta^{(1)}(\vk') \rangle'$ the linear power spectrum, and 
$P_{22}(k) = \langle \delta^{(2)}(\vk)\delta^{(2)}(\vk') \rangle'$ and $P_{13}(k) = 2 \langle \delta^{(1)}(\vk)\delta^{(3)}(\vk') \rangle'$ 
the pure loop contributions. Using eq.~(\ref{deltan}) $P_{22}$ is given by
\begin{align}
 P_{22}(k) &\equiv 2 \int \Dk{p} (F_2(\vk - \vp,\vp))^2 P_L(|\vk-\vp|) P_L(p) \\
 &= \frac{1}{2} \int \Dk{p} (k_i L_i^{(2)}(\vk - \vp,\vp))^2  P_L(|\vk-\vp|)P_L(\vp) \\  
 &+ k_i k_j k_k \int \Dk{p} L_i^{(2)}(\vk - \vp,\vp)L_j^{(1)}(\vk - \vp)L_k^{(1)}(\vp) P_L(|\vk-\vp|)P_L(\vp) \\
 &+ \frac{1}{2} \int \Dk{p} (k_i k_j  L_i^{(1)}(\vk - \vp)L_j^{(1)}(\vp))^2 P_L(|\vk-\vp|)P_L(\vp),\label{P22}
 \end{align}
where in the second equality we make use of eq.~(\ref{LPTtoF2}). Equivalently, using eqs.~(\ref{deltan}, \ref{LPTtoF3}) we have
\begin{align} 
 P_{13}(k) &\equiv  6 P_L(k) \int \Dk{p} F_3(\vk, \vp,-\vp)  P_L(p)  \nonumber\\
 &=P_L(k) \int\Dk{p} k_i L^{(3)}_i(\vk, \vp,-\vp)   P_L(p) \nonumber\\ 
 &+  2 P_L(k)\int \Dk{p}  k_ik_j L^{(2)}_i(\vk,\vp)L^{(1)}_j(-\vp)  P_L(p) \nonumber\\
 &+ P_L(k)\int \Dk{p}   k_ik_jk_kL^{(1)}_i(\vk)L^{(1)}_j(\vp)L^{(1)}_k(-\vp)   P_L(p). \label{P13}
\end{align}
Now, we use the definitions of $Q(k)$ and $R(k)$ functions\footnote{These $k$-{\it functions} were introduced in \cite{Matsubara:2007wj} 
and generalized to MG in \cite{Aviles:2017aor}.} 
in appendix \ref{app:kfunctions} to obtain
\begin{align}
 P_{22}(k) &= \frac{9}{98} Q_1(k) + \frac{3}{7} Q_2(k) + \frac{1}{2} Q_3(k) \label{RelP22RQ}\\
 P_{13}(k) &= \frac{10}{21} R_1(k) + \frac{6}{7} R_2(k) - \sigma^2_L k^2 P_L(k) \label{RelP13RQ}
\end{align}
with
\begin{equation}\label{sigma2L}
  \sigma^2_L =    \frac{1}{3}\delta_{ij} \langle \Psi_i(0)\Psi_j(0) \rangle = \frac{1}{6\pi^2}\int dp P_L(p),
 \end{equation}
the 1D variance of the of Lagrangian displacements.

In ref.~\cite{Aviles:2017aor}, the matter power spectrum constructed from eqs.~(\ref{RelP22RQ}, \ref{RelP13RQ}) 
was denoted as $P^\text{SPT*}$ due to its resemblance to 
the SPT power spectrum, and to its well known equivalence for the EdS case \cite{Matsubara:2007wj}. 
Now, we are showing that this identification holds
generally, demonstrating that $P^\text{SPT*}=P^\text{SPT}$ for generalized cosmologies. 

%
\begin{figure}
	\begin{center}
	\includegraphics[width=3 in]{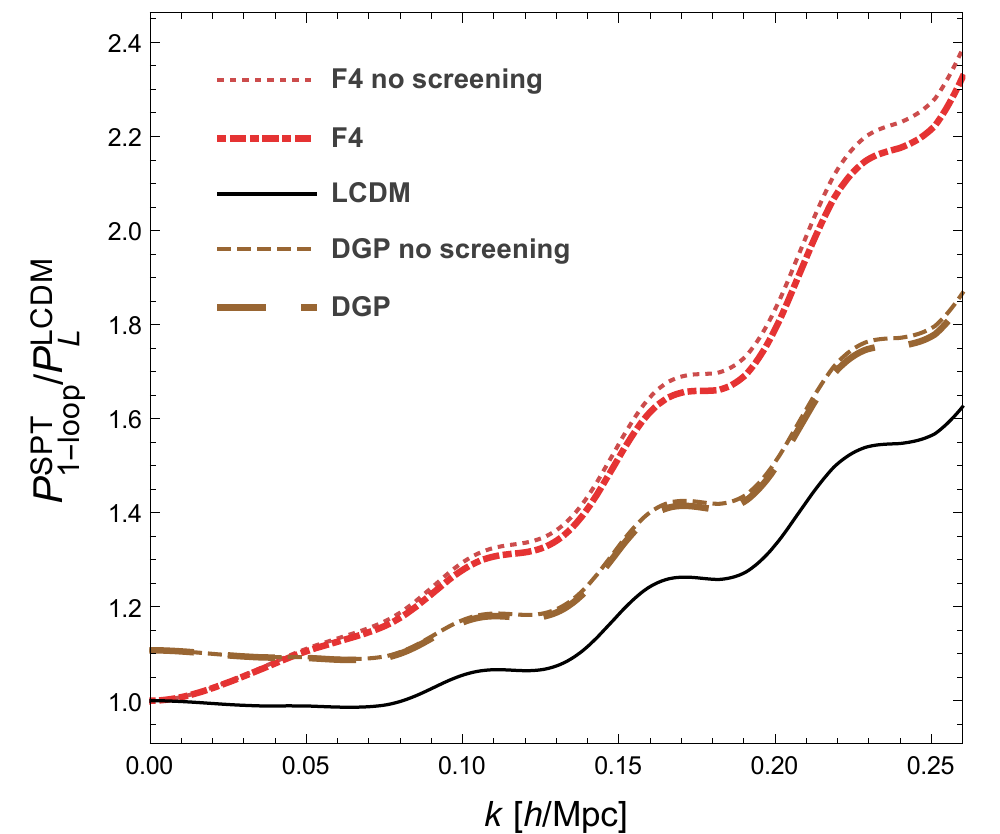}	
	\caption{ Ratio of 1-loop SPT to $\Lambda$CDM linear matter power spectra  
	 for $\Lambda$CDM, F4 and DGP models at redshift $z=0$
	 We fix cosmological parameters to the best fit of the WMAP Nine-year results \cite{Hinshaw:2012aka}.  
	\label{fig:PSmatter}}
	\end{center}
\end{figure}

In figure \ref{fig:PSmatter} we plot the ratios of the SPT 1-loop to linear matter power spectra, 
for models $\Lambda$CDM, F4 and the normal branch of DGP with crossover scale fixed by the Hubble constant, $r_c = H_0^{-1}$; 
in appendix \ref{app:mg} we give a brief summary of these models.
We fix the cosmological parameters to $\Omega_{m} = 0.281$, $\Omega_b = 0.046$, $h = 0.697$, $n_s = 0.971$,
and $\sigma_8 = 0.82$, corresponding to WMAP 9 years best fit to $\Lambda$CDM \cite{Hinshaw:2012aka}. 
For F4 model, the background cosmology is indistinguishable to that in $\Lambda$CDM. This is not the case for DGP.
However, as it is usual in the literature, 
for DGP we fix to a $\Lambda$CDM background. In such a way we can compare the differences in the growth of perturbations due to
the fifth force and not to a different Hubble flow. Despite this, the power spectrum in DGP suffers a scale independent shift 
because $A(k,t)=A(t) \neq A_0$. We plot the power spectra with and without screenings; the latter are provided by setting $M_2=M_3=0$ in eq.~(\ref{deltaI}). 
We stress out that for $\Lambda$CDM we make use of their exact kernels, that differ to those using EdS kernels by about 1\%  at quasi-linear scales.
For the three models we use the same linear $\Lambda$CDM power spectrum as input, and thereafter we rescale it with 
\begin{equation}
  P_L(k,t) = \left(\frac{D_+(k,t)}{D_+^\text{$\Lambda$CDM}(t_0)}\right)^2 P_L^{\Lambda\text{CDM}}(k,t_0),
\end{equation}
to get the linear power spectra in MG.
This prescription is valid for models sharing an early EdS phase, as the majority of MG models considered in cosmology, 
and particularly to those used here. 
Otherwise, the linear power spectrum can be computed from an Einstein-Boltzmann code as \verb|MGCAMB| \cite{Lewis:1999bs,Hojjati:2011ix}.

\end{subsection}

\end{section}

\begin{section}{Lagrangian biased tracers in modified gravity}\label{sec:3}

In large scale cosmological surveys, most of the objects of interest do not follow exactly the patterns of the underlying matter clustering, 
but their evolution is encoded in the statistics of tracers that provide biased estimations of matter statistics. 
Since we are interested in late time dynamics, where
structure formation takes place, we assume all the matter is in the form of dark matter, and baryonic effects are not accounted for; instead
all high nonlinearities are encapsulated into the bias parameters. 
In subsection \ref{sec:3.1}, we apply the bias to initial, yet linear fields, which is the Lagrangian bias approach. 
Thereafter, nonlinear evolution takes place, which is studied in subsection \ref{sec:3.2}. 

\begin{subsection}{Initially biased tracers}\label{sec:3.1}

We filter the initial matter overdensities over a spatial scale $R_{\Lambda}$, 
that smooths out small scales, $q\lesssim R_{\Lambda}$ (or $k\gtrsim \Lambda = 1/R_{\Lambda}$ in Fourier space),
nonlinear fluctuations of the overdensity field as
\begin{equation}
 \delta_R(\vq) = \int d^3q' W(|\vq-\vq'|;R_{\Lambda}) \delta_m(\vq'),
\end{equation}
and assume the existence of a function $F$ that relates the density fluctuations $\delta_X(\vq)$ of tracers  
with a given set of operators constructed out 
of the fields that enter the theory. We consider smoothed overdensities of the 
underlying dark matter field, and the additional gravitational scalar field. Nevertheless, we
may note from the Klein-Gordon equation that
\begin{equation}\label{linKGeq}
\frac{k^2}{2 a^2} \varphi = (A(k) - A_0) \tilde{\delta} =  2 A_0 \beta^{2}   \frac{k^2 \tilde{\delta}}{m^2 a^2}  + \mathcal{O}\left(  \frac{k^4}{m^4 a^4} \right),
\end{equation}
where the first equality is valid at first order in field fluctuations and the second is obtained by expanding in powers of $k^2/m^2 a^2$.
Thus, the inclusion of bias expansion dependence on $\nabla^2 \varphi$ is degenerated with a bias dependence on 
$\nabla^2 \delta$, and equivalently a bias operator $\varphi$ is degenerated with an operator $\delta$. 
For that reason we will assume a Lagrangian bias function
of the form\footnote{A more formal approach starts by constructing all invariant operators out of the 
gravitational and velocity potentials, or alternatively out of the deformation tensor $\nabla_i \Psi_j$, and include only those that are relevant up to the desire order
in PT \cite{McDonald:2009dh,Assassi:2014fva,Mirbabayi:2014zca}. Nevertheless, to do it in this way, our definition of 
eq.~(\ref{LagBiasDef2}) has to be  properly extended, and although it can be adapted to include 
a bare tidal bias  \cite{Vlah:2016bcl}, for example, it is not clear to us how to renormalize it. Hence, appealing simplicity, we consider
a bias relation given by eq.~(\ref{BiasFunc}). In this sense, our biasing scheme is not exhaustive.}
\begin{equation}\label{BiasFunc}
1+\delta_X(\vq) =  F(\delta_R(\vq),\nabla^2 \delta_R(\vq)).
\end{equation}
We thus expect that our formalism accomodates better for $k$-modes smaller than the corresponding scalar field mass. 
A similar expansion to the one in eq.~(\ref{linKGeq}) can be performed to the growth equation of linear overdensities [eq.~(\ref{lineardeltaeq})]. Showing that at sufficently
large scales we can effectively describe the effects of the fifth force, and the biasing in MG models with $m\neq 0$, with higher
order derivative operators.

We now Fourier transform eq.~(\ref{BiasFunc}) over both arguments to get
\begin{equation}\label{BiasFunc2}
 1+\delta_X(\vq) = \int \frac{d^2\bf{\Lambda}}{(2\pi)^2} \tilde{F}({\bf \Lambda}) e^{i{\bf D } \cdot {\bf \Lambda}},
\end{equation}
with vectors $\mathbf{\Lambda} = (\lambda,\eta)$ and  $\mathbf{D} = (\delta_R,\nabla^2\delta_R)$.
We call to $\lambda$ and $\eta$ the spectral bias parameters corresponding to bias operators $\delta_R$ 
and $\nabla^2\delta_R$, respectively, and we define the bias parameters as \cite{Matsubara:2008wx,Aviles:2018thp}
\begin{equation}\label{LagBiasDef2}
 b_{nm} \equiv \int \frac{d^2\mathbf{\Lambda}}{(2 \pi)^2} \tilde{F}(\mathbf{\Lambda})
 e^{-\frac{1}{2}\mathbf{\Lambda}^\text{T} \mathbf{\Sigma} \mathbf{\Lambda}}(i\lambda)^n (i \eta)^m,
\end{equation}
with a covariance matrix given by $\Sigma_{11} = \langle \delta_R^2 \rangle = \sigma^2_R$,
$\Sigma_{12} = \Sigma_{21} = \langle \delta_R \nabla^2 \delta_R \rangle$, 
and $\Sigma_{22} = \langle (\nabla^2 \delta_R)^2 \rangle$. 
We identify $b_{n0}$ with the local (in matter density) Lagrangian bias of order $n$, $b_{n0}= b_n$; while
$b_{01}$ with the linear bias in the curvature, commonly written as $b_{01}=b_{\nabla^2}$. Nevertheless, we 
keep in mind that $b_{01}$ is introduced from an operator $\nabla^2 \varphi$ which turns out, fortunately, to
be degenerated with $\nabla^2 \delta$.

The local $b_{n0}$ are already renormalized in the sense of ref.~\cite{McDonald:2006mx}, i.e., the biased tracers statistics have no zero-lag correlators.
But strictly, the $b_{nm}$ biases require further renormalization in order to 
remove subleading dependences on the smoothing kernel: For scales much larger than the smoothing $R_\Lambda$, tracers statistics should not
depend on $R_{\Lambda}$, but the existence of the BAO bump in the correlation function at about $r_\text{BAO} \simeq 110 \, \text{Mpc}/h$ with 
a width of $\Delta r_\text{BAO} \simeq 20 \, \text{Mpc}/h$ makes the smoothing-independence condition more restrictive. If it is not 
accomplished, an artificial damping of the BAO peak is produced.
This observation led to the introduction and renormalization of curvature and higher order derivative bias operators \cite{Schmidt:2012ys}. 
Under our bias definition, that renormalization is provided by replacing the factor 
$(i \eta)^n$ by $(i (\eta + W_1 R_\Lambda^2 \lambda))^{n}$ in eq.~(\ref{LagBiasDef2}),
where $W_1$ is the second term in the expansion of the filtering function in Fourier space, 
$\tilde{W}(kR_\Lambda) = \sum_{n=0}^\infty W_n (k R_\Lambda)^{2n}$ \cite{Aviles:2018thp}.  This reintroduction of the filtering kernel
makes the results independent of the smoothing scale up 
to order $\mathcal{O}(R^4_\Lambda \nabla^4 \delta_R )$, that can be further improved by considering 
higher derivatives operators as arguments of the bias function $F$. 

Leaving aside these complications, we can deal directly
with the definition (\ref{LagBiasDef2}), and keep in mind that the renormalization procedure makes the statistics $R_\Lambda$ independent.
For example, we can replace $\xi_R(q) = \langle \delta_R(\vq_1)\delta_R(\vq_2) \rangle$ by 
$\xi_L(q) =\langle \delta(\vq_1)\delta(\vq_2) \rangle$ as long as $q \gg R_\Lambda$, the other terms in the bias expansion will deal with
the fact that $\xi_R = \xi_L + 2 W_1 R_\Lambda^2 \nabla^2 \xi_L + \mathcal{O}(R_\Lambda^4 \nabla^4 \xi_L)$.


Now, we come back to eq.(\ref{BiasFunc2}) and multiply the integrand by 
$1=e^{-\frac{1}{2}\mathbf{\Lambda}^\text{T} \mathbf{\Sigma} \mathbf{\Lambda}}e^{\frac{1}{2}\mathbf{\Lambda}^\text{T} \mathbf{\Sigma} \mathbf{\Lambda}}$. We then
expand $e^{\frac{1}{2}\mathbf{\Lambda}^\text{T} \mathbf{\Sigma} \mathbf{\Lambda} + i {\bf \Lambda} \cdot {\bf D} }$  
in powers of $\lambda$ and $\eta$, and with the use of eq.~(\ref{LagBiasDef2}) we
obtain the tracer overdensity
\begin{equation}
 \delta_X(\vq) = b_{10} \delta + b_{01} \nabla^2 \delta + \frac{1}{2} b_{20} \delta^2 
 + b_{11} \delta \nabla^2 \delta + \frac{1}{2} b_{02} (\nabla^2\delta)^2 + \cdots,
\end{equation}
up to second order. Given our discussion above, we have omitted to write the label $R$ in the matter fluctuations and assume this equation
is valid for $q \gg R_{\Lambda}$. 
The linear correlation function is given by
$\xi_{X,L} = \langle \delta_X(\vq'+\vq)\delta_X(\vq')\rangle$, or
\begin{equation}
 \xi_X(q) = b_{10}^2 \xi_L(q) + 2b_{10} b_{01} \nabla^2\xi_L(q) + b_{01}^2  \nabla^4\xi_L(q),
\end{equation}
since
\begin{align}\label{linearCF}
 \nabla^2\xi_L(q) &= \langle \delta(\vq') \nabla^2 \delta(\vq'+\vq) \rangle  = -\int \Dk{k} e^{i\vk\cdot \vq} k^2 P_L(k), \\
  \nabla^4\xi_L(q) &= \langle \nabla^2  \delta(\vq') \nabla^2 \delta(\vq'+\vq) \rangle  = \int \Dk{k} e^{i\vk\cdot \vq} k^4 P_L(k). 
\end{align}
The Fourier transform gives the linear power spectrum for tracers
\begin{align}\label{linearPS}
 P_{X,L}(k) = (b_{10} - b_{01}k^2)^2 P_L(k),
\end{align}
which has the same form as the peak model (PM) linear bias with 
\begin{align}
b_{10} =  b_{10}^\text{PM}, \qquad   b_{01} = -b_{01}^\text{PM}.  
\end{align}
We notice that the notation 
$b_{nm}$ for bias is adopted also in the context of non-Gaussian linear fields \cite{Giannantonio:2009ak}, and is not related with 
our notation.

We have not assumed the time at which the initial fields are defined, and therefore the 
linear statistics we derived apply equally well to the case of Eulerian biased tracers.
If fields are evolved nonlinearly, Eulerian and Lagrangian biasing schemes give different results. 

 \end{subsection}

\begin{subsection}{Evolution of biased tracers in PT}\label{sec:3.2}

As is common practice, we make the assumption that the number of tracers is conserved,
\begin{equation}\label{tracerscons}
 (1+\delta_X(\vx))d^3x= (1+\delta_X(\vq))d^3q, 
\end{equation}
that is clearly a simplification since tracers can merge or be created. 
Developing this equation, we obtain 
\begin{align} \label{eq313}
 1+\delta_X(\vx) &= \int d^3q \, \delta_\text{D}(\vx - \vq - \Ps) (1+\delta_X(\vq))  \nonumber\\
  &= 
  \int \Dk{k} \int d^3 q   e^{i \vk \cdot (\vx -\vq -\Ps) } \int \frac{d^2 \mathbf{\Lambda}}{(2\pi)^2} 
  \tilde{F}(\mathbf{\Lambda}) e^{i \mathbf{\Lambda}\cdot \mathbf{D}},
\end{align}
where in the second equality we used eq.~(\ref{BiasFunc2}). 

We notice that the Lagrangian displacement entering eq.~(\ref{eq313}) is that of dark matter and not that of tracers. On top of that, the biasing
 is applied by using eq.~(\ref{BiasFunc}). However, tracers are subject
 to tidal forces not experienced by dark matter particles, inducing a velocity bias. Indeed, it is known that this bias
 should enter in the description as a higher derivative operator, on the same footing as the operator $\nabla^2\delta$ \cite{Desjacques:2016bnm,Mirbabayi:2014zca}. In peaks theory this bias is
 completely determined once the linear power spectrum is given and a filtering kernel is chosen \cite{Desjacques:2008jj,Desjacques:2010gz}, and shifts the Lagrangian displacement field by a term $\propto k^2 \Psi(\mathbf{k})$. 
 Having this in mind, we can think of considering
 a bias operator $\nabla \cdot \nabla^2 \Ps$ with bias parameter $b_{X,\Psi}$ as an argument of $F$. We note however that at leading order in PT it becomes
 degenerated with the density Laplacian bias operator because $\nabla \cdot \Ps^{(1)} = - \delta_L$. 
 Hence, the effect of the velocity bias, introduced here through the Lagrangian displacement, is to shift the bias parameter $b_{01}$ to $b_{01} - b_{\Psi,X}$. 
 This impact of a linear velocity bias over the biased overdensities is consistent with known results in the, more common, Eulerian approach and in the Lagrangian peaks formalism; 
 see \cite{Mirbabayi:2014zca,Baldauf:2014fza,Desjacques:2010gz} and sects.~2.7 and 6.9 of \cite{Desjacques:2016bnm}.

Now, the LPT tracers power spectrum is given by \cite{Taylor:1996ne,Matsubara:2008wx,Aviles:2018thp}
\begin{equation}\label{PX2}
 (2\pi)^3 \delta_\text{D}(\vk) + P_X^\text{LPT}(k) = \int d^3 q e^{i \vk \cdot \vq }\int \frac{d^2\mathbf{\Lambda}_1}{(2\pi)^2} 
 \frac{d^2\mathbf{\Lambda}_2}{(2\pi)^2} \tilde{F}(\mathbf{\Lambda}_1) \tilde{F}(\mathbf{\Lambda}_2) 
 \langle e^{i[\mathbf{\Lambda}_1 \cdot \mathbf{D}_1 +\mathbf{\Lambda}_2 \cdot \mathbf{D}_2  +  \vk\cdot \Delta]}\rangle,
\end{equation}
where $\mathbf{D}_{1,2} = (\delta_R(\vq_{1,2}),\nabla^2 \delta_R(\vq_{1,2}))$ and 
\begin{align}\label{LPTPS1}
 \Delta_i = \Psi_i(q_2) - \Psi_i(q_1) &= \int \Dk{k} (e^{i \vk \cdot \vq_2}-e^{i \vk \cdot \vq_1}) \Psi_i(\vk) \nonumber\\ 
 &=\sum_{n=1}^{\infty} \int \Dk{k} (e^{i \vk \cdot \vq_2}-e^{i \vk \cdot \vq_1}) \Psi_i^{(n)}(\vk)
\end{align}
is the difference of Lagrangian displacement fields separated by a distance $\vq = \vq_2 - \vq_1$. We will employ 
the cumulant expansion theorem
\begin{equation}
\langle e^{i X} \rangle = \exp \left( \sum_{N=1}^\infty \frac{i^N}{N!} \langle X^N \rangle_c \right)
  = \exp \left( -\frac{1}{2} \langle X^2 \rangle_c  -\frac{i}{6} \langle X^3 \rangle_c + \cdots \right), 
\end{equation}
where $\langle X^N \rangle_c$ denotes the cumulant of $X^N$.  In this case we have 
$X=\mathbf{\Lambda}_1 \cdot \mathbf{D}_1 +\mathbf{\Lambda}_2 \cdot \mathbf{D}_2  +  \vk\cdot \Delta$.

The strategy now is to expand all terms that contain bias spectral parameters $\lambda$ and $\eta$ with the exception of a term
$e^{-\frac{1}{2}\mathbf{\Lambda}^\text{T} \mathbf{\Sigma} \mathbf{\Lambda}}$, and use the definition of eq.~(\ref{LagBiasDef2}) to
obtain the biases.
Keeping the local bias up to second order and the curvature bias to first order we obtain
\begin{align}\label{XLPTPS}
 (2\pi)^3 \delta_\text{D}(\vk) &+ P_X^\text{LPT}(k) = \int d^3 q e^{i \vk \cdot \vq } e^{ -\frac{1}{2}k_ik_j A_{ij} - \frac{i}{6} k_ik_jk_k W_{ijk} }
   \Bigg[ 1  + b_{10}^2 \xi_L + 2 i b_{10} k_i U_i + \frac{1}{2} b_{20}^2 \xi_L^2   \nonumber\\
  &-(b_{20}  + b_{10}^2) k_i k_j U_i U_j + 2i b_{10} b_{20} \xi_L k_i U_i 
  + i b_{10}^2 k_i U_i^{11} + i b_{20} k_i U_i^{20}- b_{10} k_ik_j A_{ij}^{10} \nonumber\\
  &+2 b_{10} b_{01} \nabla^2\xi_L - 2 i b_{01} k_i \nabla_i \xi_L 
   +b_{01}^2 \nabla^4 \xi_L \Bigg],
\end{align}
with functions \cite{Carlson:2012bu}
\begin{align}
 U^{mn}_i(\vq)&=\langle \delta^{m}_L(\vq_1)\delta^{n}_L(\vq_2) \Delta_i \rangle_c,  \label{Uimn}\\
  A_{ij}^{mn}(\vq) &= \langle  \delta^{m}_L(\vq_1)\delta^{n}_L(\vq_2)  \Delta_i \Delta_j \rangle_c, \label{Aijmn}\\ 
  W_{ijk}(\vq) &= \langle \Delta_i \Delta_j \Delta_k \rangle_c, \label{Wijk}
\end{align}
and $U_i=U^{10}_i$, $A_{ij}=A_{ij}^{00}$. In appendix \ref{app:qfunctions} we find the explicit expressions for these \emph{q-functions} in
generalized cosmologies.
The complete expansion that includes second order curvature bias can be found in \cite{Aviles:2018thp}; the terms we are neglecting here are subdominant. 
One can continue this process to include higher order biases, but this necessarily needs the introduction of higher order fluctuations.

If we keep all terms quadratic in $k$ in the exponential of eq.~(\ref{XLPTPS}) and expand the rest, we can Fourier transform it and, by performing 
several multivariate Gaussian integrations, obtain an analytical expression for the correlation function. This is the approach of CLPT, first developed in
\cite{Carlson:2012bu}. But, in order to treat on an equal footing linear and nonlinear fields contributions, 
in this work we keep exponentiated only the linear piece of $A_{ij}$ and expand the rest, as was done in \cite{Vlah:2015sea}, 
obtaining the CLPT correlation function for tracers,
\begin{align}\label{XCLPTCF}
1 + \xi_X^\text{CLPT}(r) &= 1 + \xi^\text{CLPT}(r) + b_{10} \text{x}_{10}(r)+ b_{10}^2 \text{x}_{20}(r)+ b_{20} \text{x}_{01}(r) 
+ b_{10} b_{20} \text{x}_{11}(r) \nonumber\\
&\quad + b_{01}^2 \text{x}_{02}(r)+ 2(1+b_{10})b_{01} \text{x}_{\nabla^2}(r) + b_{01}^2 \text{x}_{\nabla^4}(r), 
\end{align}
with matter correlation function
\begin{align}\label{mCLPTCF}
1 + \xi^\text{CLPT}(r) &= \int  \frac{d^3 q}{(2 \pi)^{3/2} |\mathbf{A}_L|^{1/2}} e^{- \frac{1}{2}(\ve r-\vq)^\mathbf{T}\mathbf{A}_L^{-1}(\ve r-\vq) }
\Bigg( 1 - \frac{1}{2} A_{ij}^{loop}G_{ij} +\frac{1}{6}\Gamma_{ijk}W_{ijk} \Bigg), 
\end{align}
where 
$g_{i}= (A_L^{-1})_{ij} (q_j - r_j)$, $G_{ij} = (A_L^{-1})_{ij} - g_i g_j$, and $ \Gamma_{ijk} = (A_L^{-1})_{\{ij}g_{k\}} + g_i g_j g_k$.
The ``1'' in between the parentheses corresponds to the Zel'dovich approximation.
The notation $\text{x}_\text{NM}$ (and for the $\text{a}_\text{NM}$ introduced below) means that these bias 
contributions are multiplied by the linear local bias to the $\text{N}$ power times
the second order local bias to the $\text{M}$ power. These functions are 
\begin{align}
 \text{x}_{10}(r) &=  \int  \frac{d^3 q}{(2 \pi)^{3/2} |\mathbf{A}_L|^{1/2}} e^{- \frac{1}{2}(\ve r-\vq)^\mathbf{T}\mathbf{A}_L^{-1}(\ve r-\vq) }  
                            (-2 U_i g_i - A^{10}_{ij}G_{ij}), \\
 \text{x}_{20}(r) &=  \int  \frac{d^3 q}{(2 \pi)^{3/2} |\mathbf{A}_L|^{1/2}} e^{- \frac{1}{2}(\ve r-\vq)^\mathbf{T}\mathbf{A}_L^{-1}(\ve r-\vq) }   
                             (\xi_L - U_iU_jG_{ij}- U_i^{11}g_i),\\
 \text{x}_{01}(r) &=  \int  \frac{d^3 q}{(2 \pi)^{3/2} |\mathbf{A}_L|^{1/2}} e^{- \frac{1}{2}(\ve r-\vq)^\mathbf{T}\mathbf{A}_L^{-1}(\ve r-\vq) }    
                             (-U_i^{20}g_i - U_iU_jG_{ij}),  \\
 \text{x}_{11}(r) &=  \int  \frac{d^3 q}{(2 \pi)^{3/2} |\mathbf{A}_L|^{1/2}} e^{- \frac{1}{2}(\ve r-\vq)^\mathbf{T}\mathbf{A}_L^{-1}(\ve r-\vq) }   
                               (- 2 \xi_L U_i g_i), \\
 \text{x}_{02}(r) &=  \int  \frac{d^3 q}{(2 \pi)^{3/2} |\mathbf{A}_L|^{1/2}} e^{- \frac{1}{2}(\ve r-\vq)^\mathbf{T}\mathbf{A}_L^{-1}(\ve r-\vq) }    
                              \frac{1}{2} \xi_L^2,\\
 \text{x}_{\nabla^2}(r) &=  \int \frac{d^3 q}{(2 \pi)^{3/2} |\mathbf{A}_L|^{1/2}} e^{- \frac{1}{2}(\ve r-\vq)^\mathbf{T}\mathbf{A}_L^{-1}(\ve r-\vq) }   
                               \nabla^2 \xi_L(q), \\
 \text{x}_{\nabla^4}(r) &=  \int \frac{d^3 q}{(2 \pi)^{3/2} |\mathbf{A}_L|^{1/2}} e^{- \frac{1}{2}(\ve r-\vq)^\mathbf{T}\mathbf{A}_L^{-1}(\ve r-\vq) } 
                                \nabla^4 \xi_L(q).  \label{n4xi}
\end{align}

The explicit computation that leads to eq.~(\ref{XCLPTCF}), yields also a 
term $\nabla_i \xi_L(q) g_i$. However, as noted in \cite{Vlah:2016bcl}, this is highly degenerated with $\nabla^2 \xi_L(q)$: indeed, 
both terms correspond to a contribution $-k^2 P_L$ to the SPT power spectrum. 
Therefore, we have substituted the combination $2 b_{10} b_{01} \nabla^2 \xi_L(q) + 2  b_{01} \nabla_i \xi_L(q) g_i$
by $2 (1+b_{10}) b_{01} \nabla^2 \xi_L(q)$ when writing eq.~(\ref{XCLPTCF}).

\bigskip

LPT describes quite well the BAO wiggles in the power spectrum, but it fails to follow its broadband trend. Since here we 
are interested in the latter also, we expand all terms out of the exponential in eq.~(\ref{XLPTPS}) and perform the $\vq$ integration to obtain the  
SPT power spectrum for Lagrangian biased tracers. 
This procedure involves lengthy mathematical manipulations 
that are presented in appendix \ref{app:PS}, yielding
\begin{align}\label{XSPTPS}
  P^\text{SPT}_X(k)  &=  P_L(k)  + P_{22}(k) + P_{13}(k) + b_{10} \text{a}_{10}(k) + b_{20} \text{a}_{01}(k)  + b_{10}^2 \text{a}_{20}(k)  \nonumber\\
  & \quad + b_{10} b_{20} \text{a}_{11}(k) + b_{20}^2 \text{a}_{02}(k)
     - 2(1+b_{10}) b_{01} k^2 P_L(k) + b_{01}^2 k^4 P_L(k) 
\end{align}
with $P_{22}$ and $P_{13}$ given by eqs.~(\ref{RelP22RQ}, \ref{RelP13RQ}), and
\begin{align}
  \text{a}_{10}(k) &= 2 P_L(k) + \frac{10}{21}R_1(k) +\frac{6}{7} R_{1+2}(k) +\frac{6}{7} R_{2}(k) + \frac{6}{7} Q_{5}(k) + 2 Q_7(k)- 2\sigma^2_L k^2 P_L(k),   \\
  \text{a}_{01}(k) &=  Q_{9}(k) + \frac{3}{7}Q_8(k), \\
  \text{a}_{20}(k) &= P_L(k)  + \frac{6}{7}R_{1+2}(k)  + Q_9(k)+ Q_{11}(k)- \sigma^2_L k^2 P_L(k),\\
  \text{a}_{11}(k) &= 2 Q_{12}(k), \\ 
  \text{a}_{02}(k) &= \frac{1}{2} Q_{13}(k),
\end{align} 
where $Q$ and $R$ functions are given in appendix \ref{app:kfunctions}. It is worth to pointing out that the results of Sect.~\ref{sec:2.2} are allowing us to 
identify eq.~(\ref{XSPTPS}) with the SPT power spectrum with Lagrangian biased tracers in generalized cosmologies.
We further notice that its structure is similar to that of \cite{Matsubara:2008wx}; but
in MG the $Q$ and $R$ functions are different.  Also, here we find a function $R_{1+2}$, 
which for EdS coincides with the combination $R_1 + R_2$.

\end{subsection}

\begin{subsection}{Numerical results}\label{sec:3.3}

\begin{figure}
	\begin{center}
	\includegraphics[width=3 in]{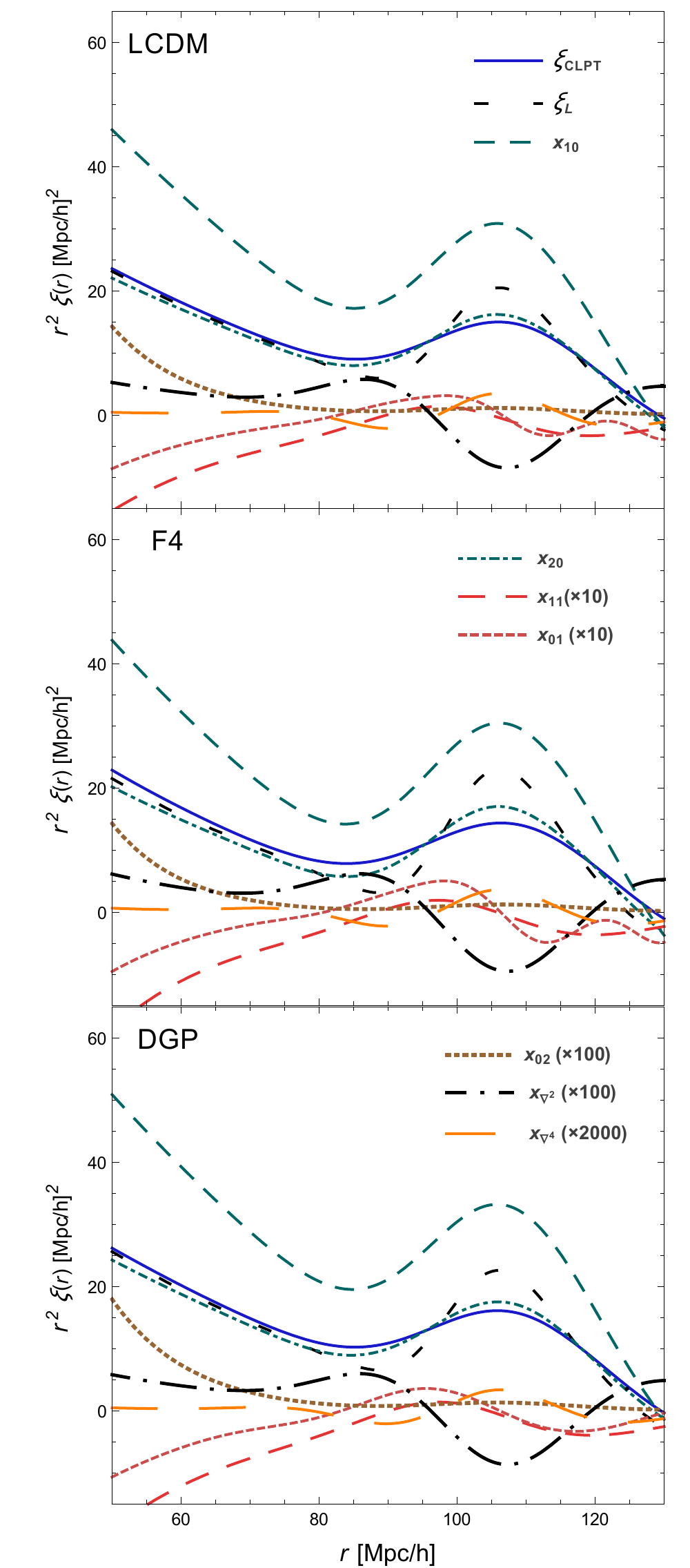}	
	\includegraphics[width=2.97 in]{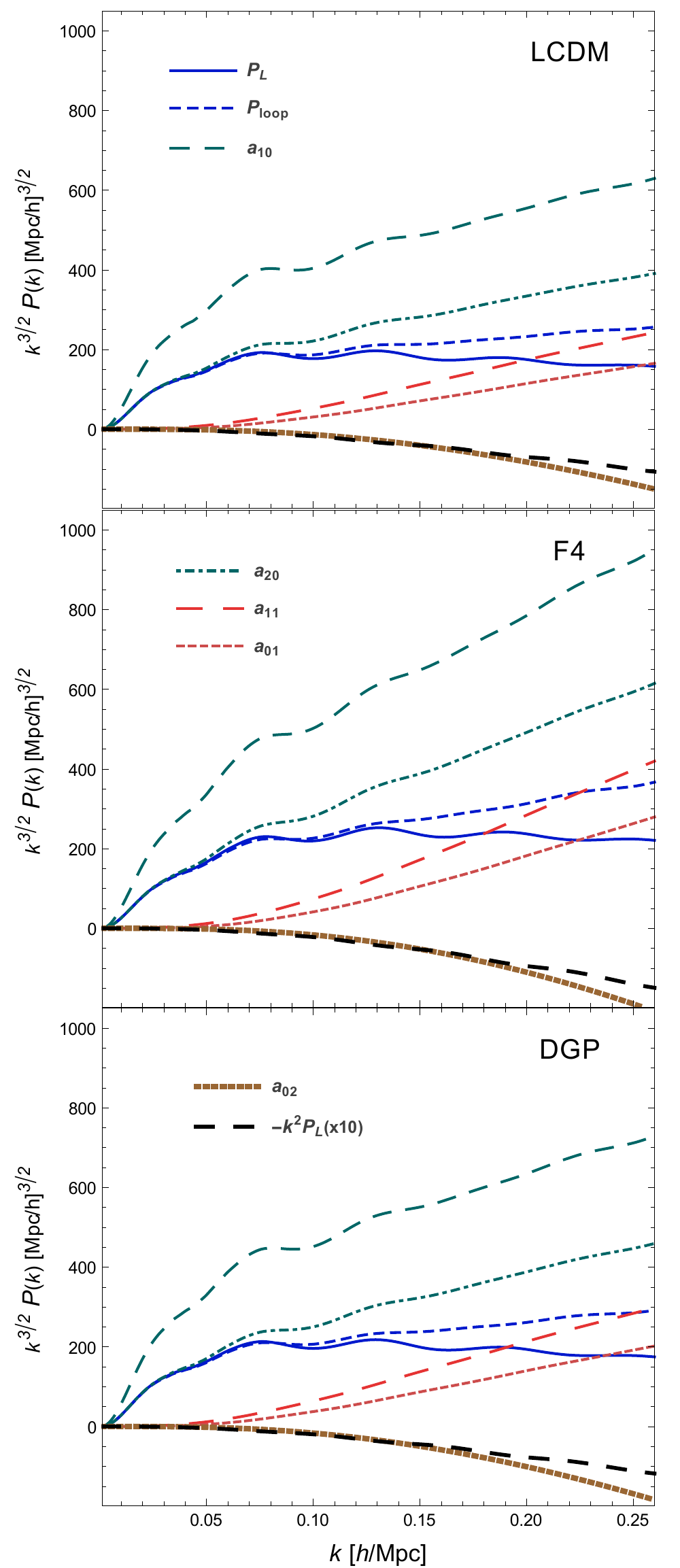}
	\caption{Bias components for the CLPT correlation function [eq.~(\ref{XCLPTCF})] and SPT power spectrum [eq.~({\ref{XSPTPS}})]
	for models $\Lambda$CDM model, F4 and the normal branch of DGP with crossover scale $r_c=1/H_0$.
	 We fix cosmological parameters to the best fit of the WMAP Nine-year results \cite{Hinshaw:2012aka} and evaluate at redshift $z=0$. 
	\label{fig:BiasComp}}
	\end{center}
\end{figure}

\begin{figure}
	\begin{center}
	\includegraphics[width=3 in]{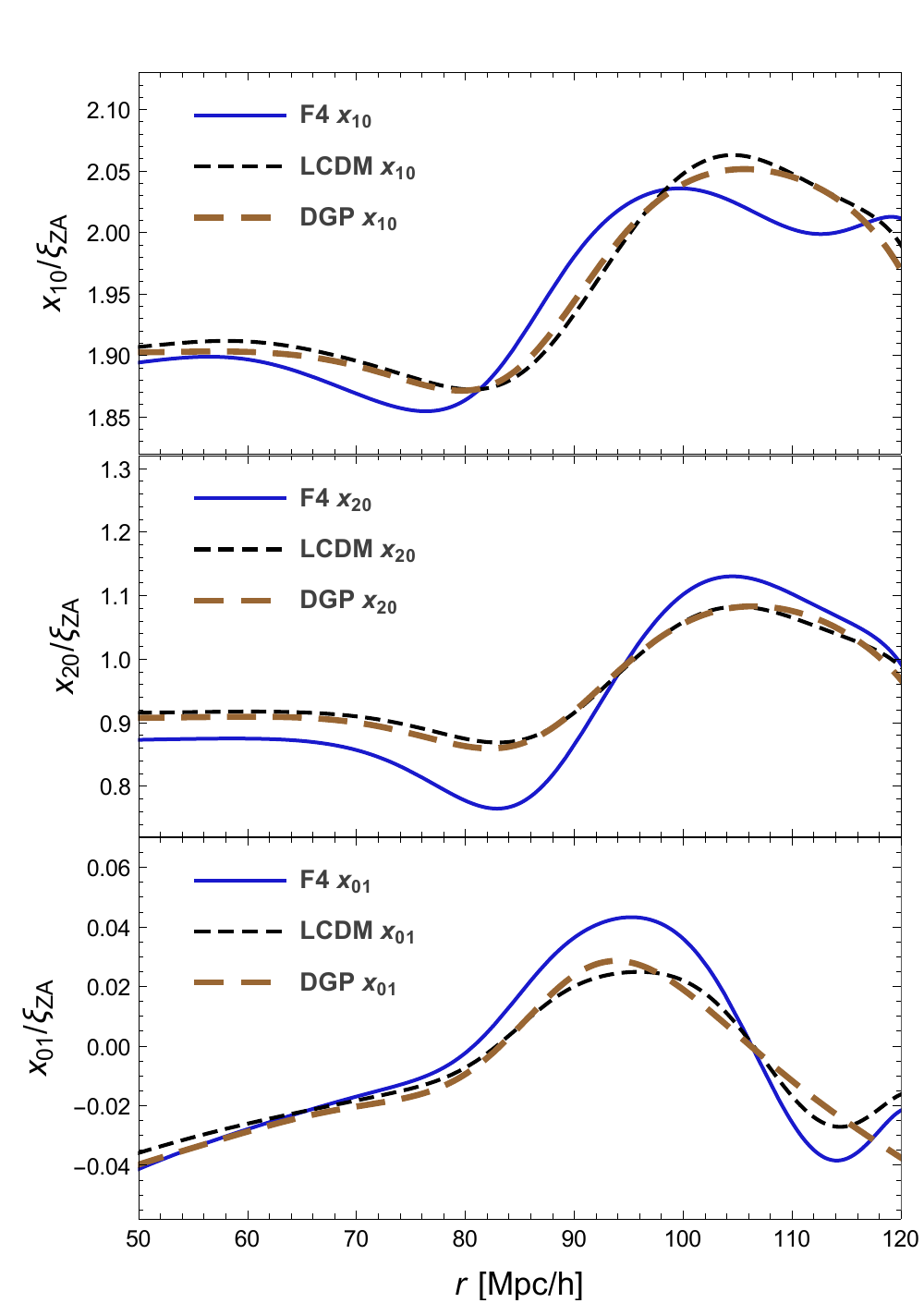}
	\includegraphics[width=3 in]{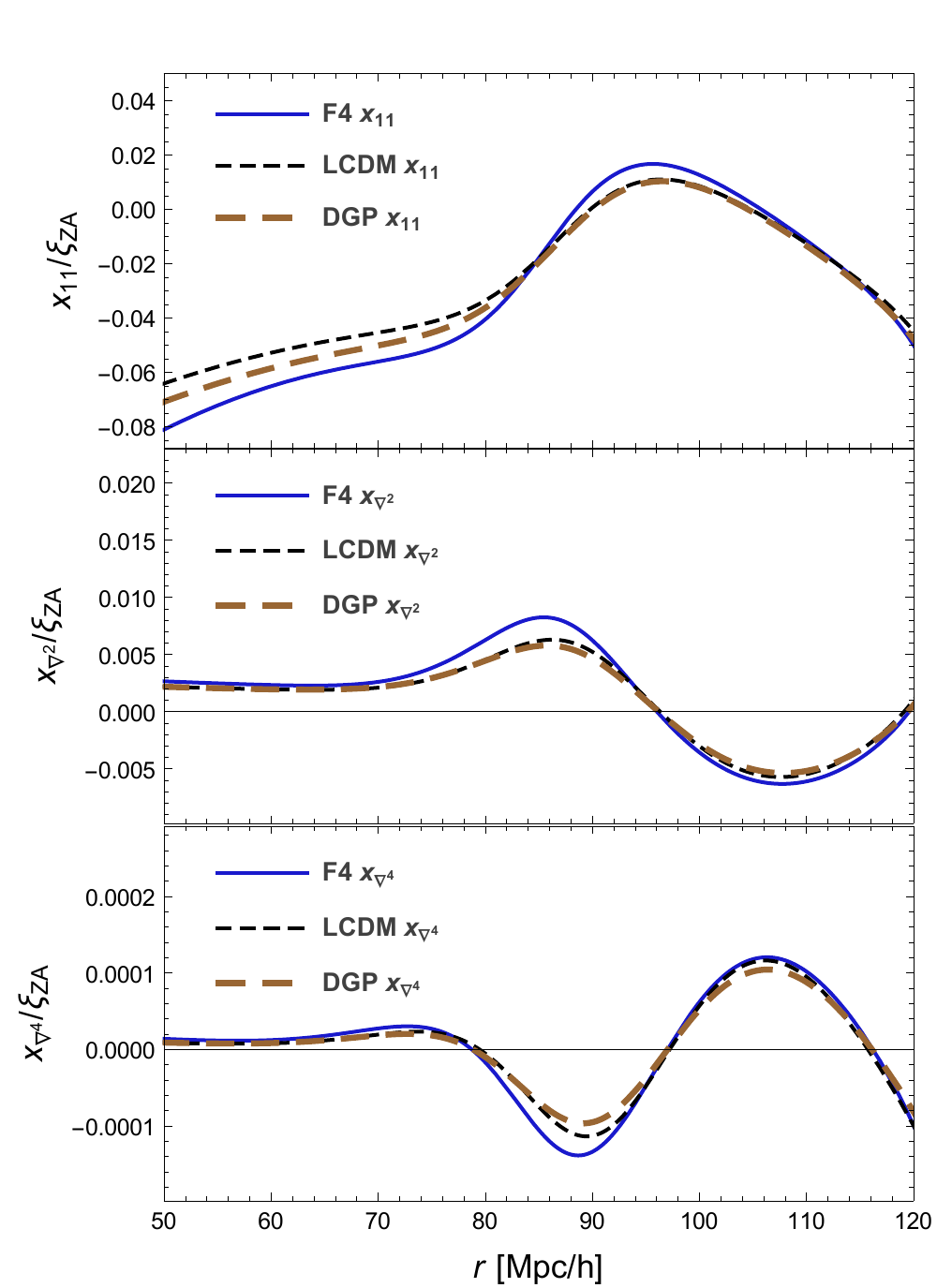}
	\caption{Ratios of bias functions $\text{x}_\text{MN}$ over the Zel'dovich approximation correlation function.
	\label{fig:CFBiasratios}}
	\end{center}
\end{figure}

To compute the SPT power spectrum and the CLPT correlation function we have developed the code \verb|MGPT| that we make public available
with this article.\footnote{The code is available at \href{www.github.com/cosmoinin/MGPT}{www.github.com/cosmoinin/MGPT}} 
First, it calculates the set of $Q$ and $R$ functions in appendix \ref{app:kfunctions}, 
for which the angular $x$ integration uses a Gauss-Legendre quadrature 
and the radial part uses the trapezoidal quadrature. At each step
of the integration we solve the differential equations eqs.~(\ref{DAeveq}, \ref{DBeveq}) to obtain the functions 
$D_\mathcal{A}$, $D_\mathcal{B}$  using the solver
\emph{bsstep} \cite{press1992}, as well as the corresponding third order growth functions. The ``time'' variable is chosen as $\eta=\ln a$, with $a$ the scale factor, starting with EdS initial conditions at $\eta_{ini}=-6$.   
Thereafter, the code computes the $q$-functions of appendix \ref{app:qfunctions} and the CLPT contributions of eqs.~(\ref{mCLPTCF}-\ref{n4xi}).

\begin{figure}
	\begin{center}
	\includegraphics[width=3 in]{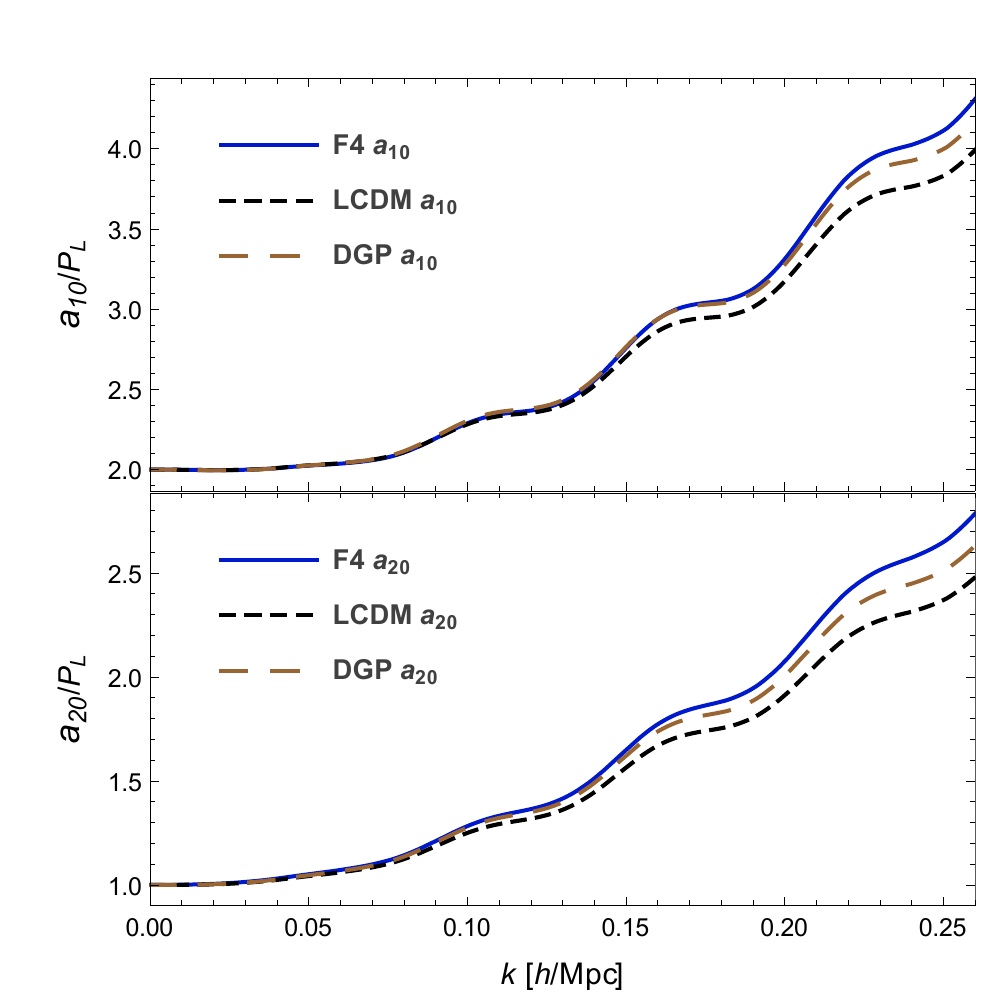}
	\includegraphics[width=3 in]{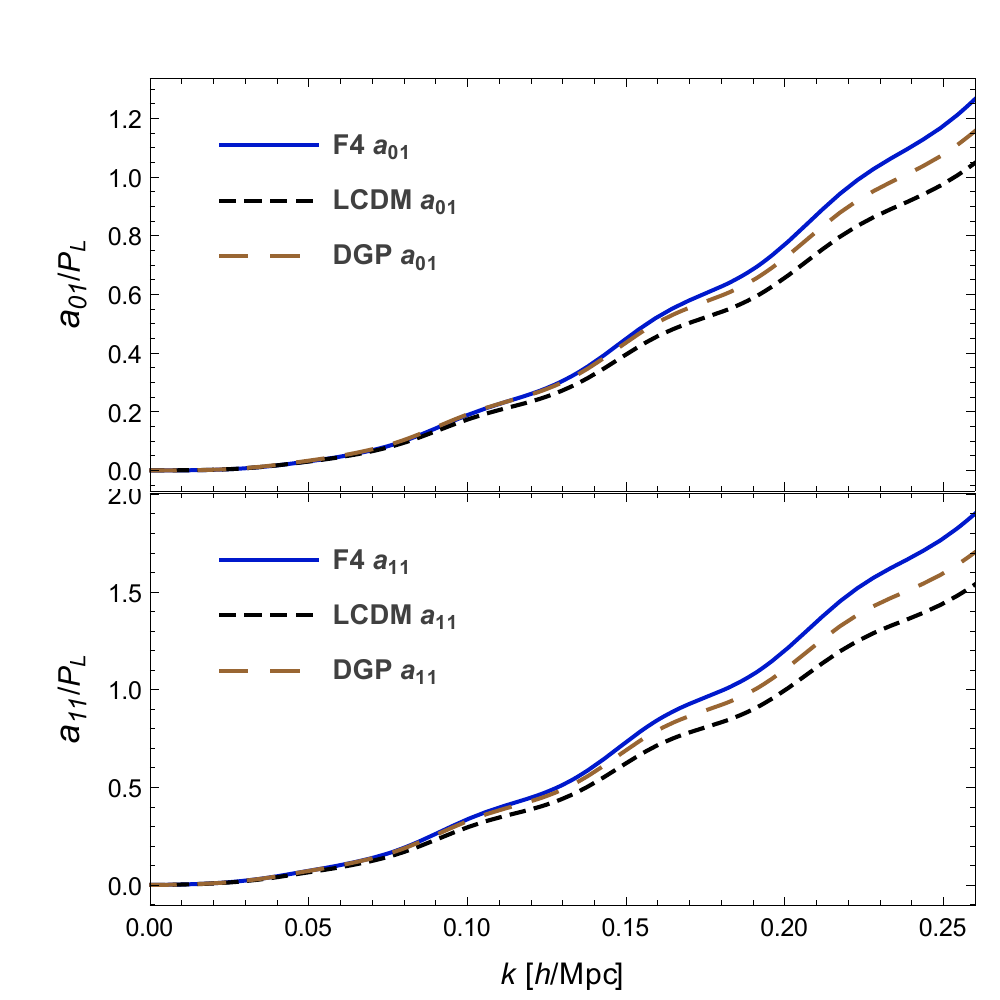}
	\caption{Ratios of bias functions $\text{a}_\text{MN}$ over the linear power spectrum.
	\label{fig:Biasratios}}
	\end{center}
\end{figure}

In figure \ref{fig:BiasComp}
we show the different contributions $a_\text{NM}$ and $\text{x}_\text{NM}$ to the tracers power spectrum and correlation 
function for models $\Lambda$CDM, F4 and  the normal branch of DGP with crossover scale $r_c=1/H_0$. Though these specific models are already ruled out 
by observations, they show the kind of growth of fluctuations expected in MG theories.
As in section \ref{sec:2.2}, in the $\Lambda$CDM case we make use of their exact kernels. 

In figure \ref{fig:CFBiasratios} we plot the ratios of the $\text{x}_\text{NM}$ bias contributions to the Zel'dovich approximation correlation function for each model; 
while in figure \ref{fig:Biasratios} we plot the ratios of the $\text{a}_\text{NM}$ bias contributions to the linear power spectrum for each model.
 
\end{subsection}

\end{section}

\begin{section}{Simple model for halo bias in $f(R)$ gravity}\label{sec:4}

So far, we were concerned in how the bias parameters appear in the structure of LPT and SPT statistics, saying nothing 
about their own evolution. In this section we put forward a model for the estimation of local bias in MG, which although is not rigorous, reflects
the following observation: Generally MG is scale dependent and because the fifth force is attractive (in general), 
matter fluctuations grow faster than in GR. This implies that 
the critical density for collapse $\delta_c$ is smaller in MG. 
On the other hand, by the same reason the power spectrum acquires more strength and the variance 
\begin{equation}\label{variance}
S(R) \equiv\sigma^2_{R} = \int \Dk{k} |\tilde{W}(k R)|^2 P_L(k)
\end{equation}
becomes larger in MG than in GR. 
Typically, a local bias depends on the critical density for collapse $\delta_c$ and on the fluctuations variance, schematically 
$b_n \sim \left(\frac{\delta_c}{\sigma_R^2}\right)^n + \cdots$,
which is a good approximation for massive halos. This implies that
\begin{equation} \label{guessBias}
 b_n^\text{MG}  <   b_n^\text{GR},  
\end{equation}
reflecting that halos are more efficiently formed in MG than in GR. 
This effect has been observed recently in N-body simulations \cite{Arnold:2018nmv}. The rest of this section is aimed to show this property
in $f(R)$ theories.

\begin{subsection}{Bias model}\label{sec:4.1}

There are some obstacles when describing bias in generalized cosmologies. 
First, the scale dependence of growth functions implies that
even linear bias is scale dependent; second, Birkhoff's theorem is not valid, and therefore cannot be applied to 
spherical collapse calculations or to the peak-background split (PBS) prescription. 
From the work of \cite{Hui:2007zh} we know that  
the Lagrangian linear local bias evolves  as $b_1(z) \propto D^{-1}_+(k,z)$. Bearing this in mind, perhaps the most easy way to get a model for bias parameters
is to compute it from a universal mass function at sufficiently large redshift $z_{i}$ at which the evolution 
is indistinguishable from GR (as is the case for the models we have used to show our results), 
and the peak background split procedure is valid. 
In such a way we obtain a bias for GR as a function of mass $b(R(M),t_{ini})$  and thereafter we evolve it 
with a growth function that we choose as\footnote{This growth function was proposed in \cite{Parfrey:2010uy} since excursion sets 
calculations are particularly sensitive to the growth of the variance of smoothed fields. }

\begin{equation} \label{defDplusM}
 \tilde{D}_{+}^\text{MG}(M,z;z_i) \equiv \frac{\sigma_{R,\text{MG}} (M,z)}{\sigma_{R,\text{MG}}(M,z_i)}
 = \frac{\sigma_{R,\text{MG}}(M,z)}{\sigma_{R,\text{$\Lambda$CDM}}(M,z)} D_+^\text{$\Lambda$CDM}(z;z_i)
 \end{equation}
where $M$ is the mass enclosed by the spherical perturbation and $R = (3 M/4\pi \bar{\rho}_m(z=0))^{1/3}$ its Lagrangian radius. 
The only requirement we adopt is that at the initial redshift $z_i$, the evolution of fluctuations in MG and GR are the same for all scales of interest. 

\begin{figure}
	\begin{center}
	\includegraphics[width=3 in]{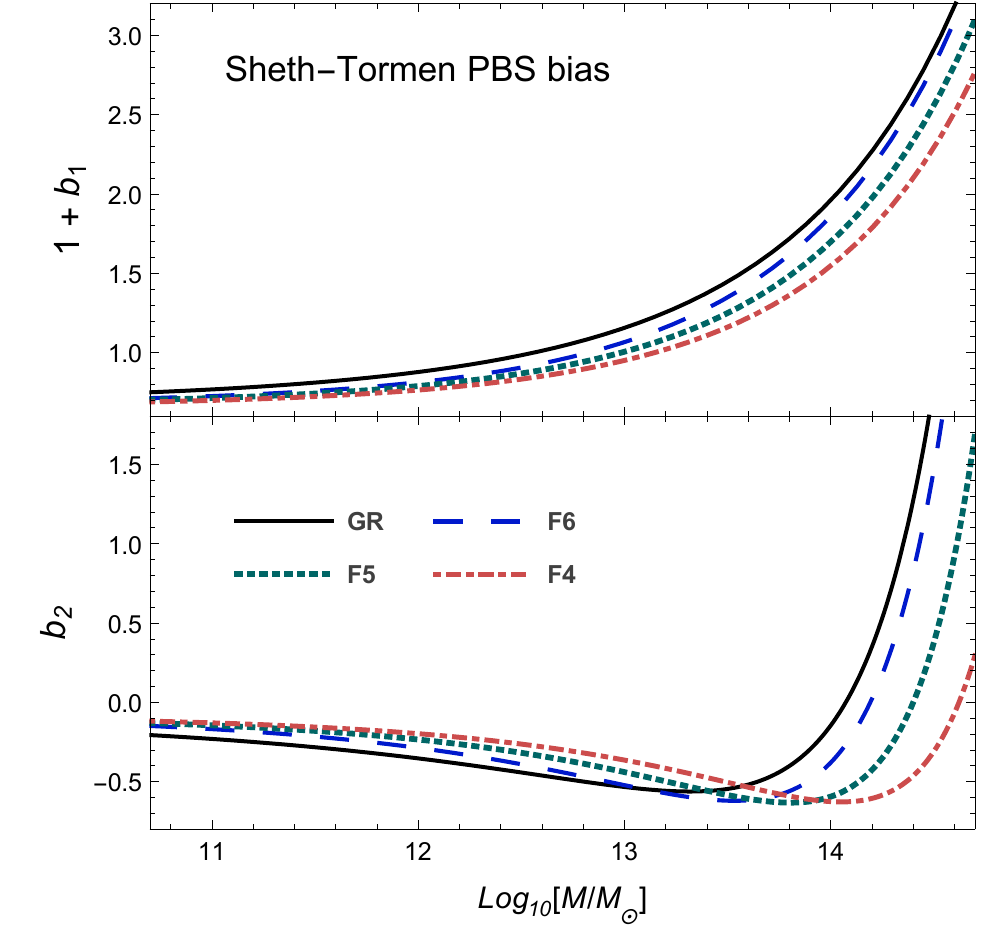}
	\caption{Analytical large scale bias $b_{LS} =1+b_1$ and bias $b_2$ for models $\Lambda$CDM, F6, F5 and F4. 
	We use eqs.~(\ref{b1}, \ref{b2}) with Sheth-Tormen parameters $q=0.707$ and $p=0.3$. The filtering function
	is a top-hat of width given by the Lagrangian radius $R(M)$.
	\label{fig:BiasST}}
	\end{center}
\end{figure}

We consider a mass function that gives the number density of halos over a mass interval $(M,M+dM)$,
\begin{equation}\label{massfunction}
n(M,z) =\frac{\bar{\rho}}{M^2} \nu f(\nu) \left| \frac{d \log \nu}{d \log M}\right|,  
\end{equation}
with $\nu f(\nu)$ the multiplicity function and 
$\nu = \delta_{c}/\sigma_{R}$ is the peak significance threshold. 
We apply the PBS prescription to obtain the local biases at redshift $z_i$
\begin{align}\label{bnOfMini}
b_n(M,z_i) &=  
\frac{(-1)^n}{\sigma^n_R(M,z_i)}\frac{1}{\nu_i f(\nu_i)}\frac{d^n \nu_i  f (\nu_i)}{d \nu_i^n}, 
\end{align} 
and thereafter we evolve them to redshift $z$ with the growth of eq.~(\ref{defDplusM}) as
\begin{align} \label{bnOfM}
b_n(M,z)&=\frac{1}{[\tilde{D}_+^\text{MG}(M,z;z_i)]^n} b_n(M,z_i) =
           \frac{(-1)^n}{\sigma^n_R(M,z)}\frac{1}{\nu_i f(\nu_i)}\frac{d^n \nu_i  f (\nu_i)}{d \nu_i^n}, 
\end{align}
with $\nu_i=\nu(z_i)$.
It is necessary to specify the time of evaluation of the peak significance
because unlike in GR, it is time dependent in MG. We emphasize that the bias parameters obtained from the PBS prescription
coincide with those defined in eq.~(\ref{LagBiasDef2}), as it was shown in \cite{Aviles:2018thp}.


In $\Lambda$CDM the density collapse threshold $\delta_c^\text{GR}$ is scale independent and weakly depends on the cosmology. Instead, in MG it 
becomes scale dependent and also dependent on the environmental density $\delta_{env}$ because of violation of Birkhoff's theorem. 
It  is possible to obtain a unique
function $\delta_c^\text{MG}(M,z)$ by averaging over environmental overdensities or by choosing an specific one
out of a known distribution of environments \cite{Li:2011qda}.
An alternative route is adopted in \cite{Kopp:2013lea}, where that average is performed to the 
initial conditions using Gaussian peaks model \cite{Bardeen:1985tr}; 
that work also provides a fitting function for the top-hat density collapse threshold in $f(R)$ gravity, 
that we will use also here to exemplify our bias model.

\begin{figure}
	\begin{center}
	\includegraphics[width=3 in]{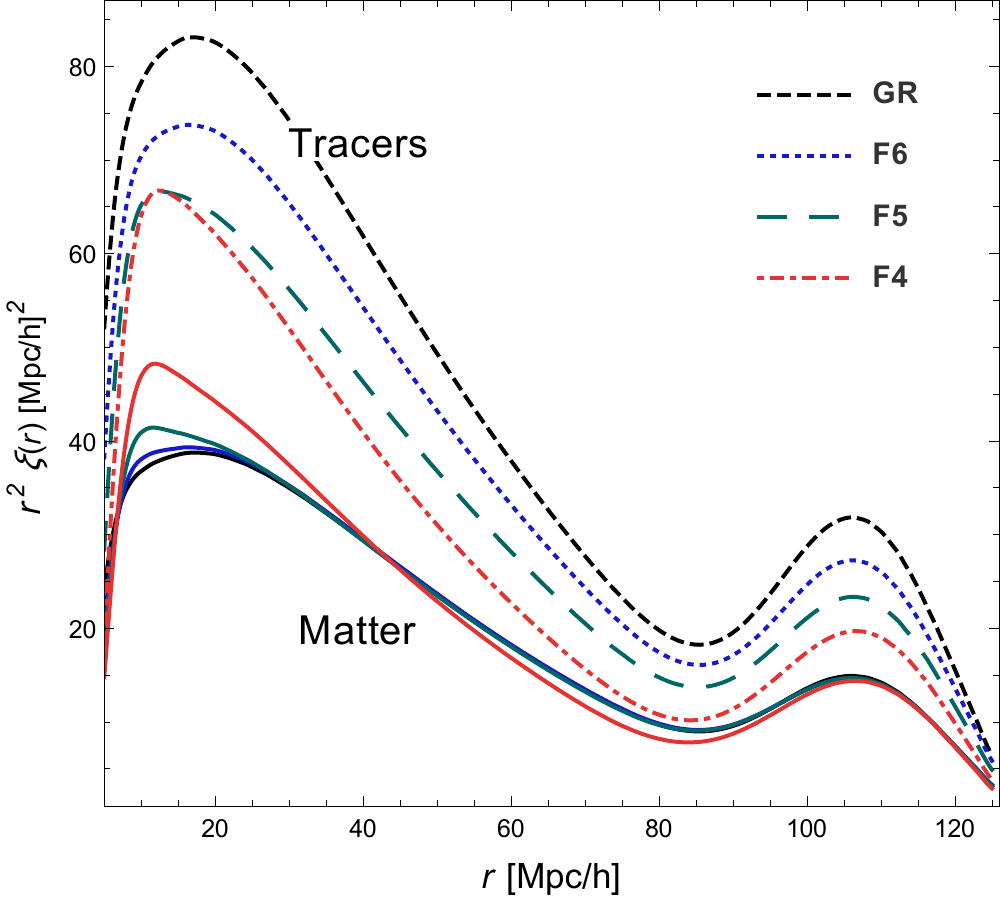}
	\caption{Correlation function for tracers (non-solid curvers) using first and second order biases from the results of figure \ref{fig:BiasST}, with 
	the halo masses
	fixed to $10^{13.5} M_\odot$. The lower, solid curves, show the correlation functions for matter. We plot the models $\Lambda$CDM (black), 
	F6 (blue), F5 (green) and F4 (red).
	\label{fig:CFbias}}
	\end{center}
\end{figure}

The multiplicity function may be parametrized as \cite{Sheth:1999mn}
\begin{equation}\label{multfunc}
 \nu f(\nu) = \mathcal{N} \sqrt{\frac{2}{\pi}q \nu^2} (1+ (q \nu^2 )^{-p} ) e^{-q \nu^2/2},
\end{equation}
where the number $\mathcal{N}$ normalizes the multiplicity as $\int_0^\infty d\nu f(\nu) = 1/2$. The two free parameters take values
$q=1$ and $p=0$ for Press-Schechter \cite{Press:1973iz}, and $q=0.707$ and $p=0.3$ for Sheth-Tormen \cite{Sheth:1999mn}
mass functions. 
The linear and second order local biases become
\begin{align}
 b_1(M) &= \frac{1}{\tilde{D}_+} \frac{1}{\delta_c}\left( q \nu^2 -1 + \frac{2 p}{1+(q \nu^2)^p} \right), \label{b1}\\
 b_2(M) &= \frac{1}{\tilde{D}_+^2} \frac{1}{\delta_c^2}\left( q^2 \nu^4 -3 q \nu^2 + \frac{2 p(2q \nu^2 + 2p -1)}{1+(q \nu^2)^p} \right) \label{b2}.
\end{align}

 It has been shown that the Sheth-Tormen model gives good 
results for the halo mass function also in $f(R)$ gravity \cite{Lombriser:2013wta}.
Motivated by this, we compute our analytical formula for biases of eqs.~(\ref{b1}-\ref{b2}) fed with the Sheth-Tormen mass
function parameters and using a top-hat filter in the variance. We plot the cases $b_1$ and $b_2$ in figure \ref{fig:BiasST} for 
GR, F6, F5 and F4 models. 
From here we strenghten our heuristic result in eq.~(\ref{guessBias}). In the next subsection we compare our bias model to
results obtained from excursion set theory.

Figure \ref{fig:CFbias} shows the CLPT correlation functions for matter and for tracers with a fixed halo mass of $10^{13.5} M_\odot$, computed with
the biases $b_1$ and $b_2$ shown in figure \ref{fig:BiasST} and using eq.~(\ref{XCLPTCF}). This shows that regardless the MG model, 
the matter correlation functions are quite similar at 
large, linear scales, while for tracers they can differ significatively.

\end{subsection}

\begin{subsection}{Excursion sets}\label{sec:ES}

Excursion set theory \cite{Bond:1990iw} identifies the places where virialized structures will form with overdensities that, when smoothed
over that same region, exceed some critical value, that we take to be the top-hat density collapse $\delta_c$. 
Letting the smoothing lenght vary from $\infty$ to $R$ makes $\delta_R(\vq)$ to describe a random walk with $R$ as the evolution
variable, although it is convenient to parametrize the trajectories in terms of the variance $S(R)$
which decrease with $R$.
\begin{figure}
	\begin{center}
	\includegraphics[width=6 in]{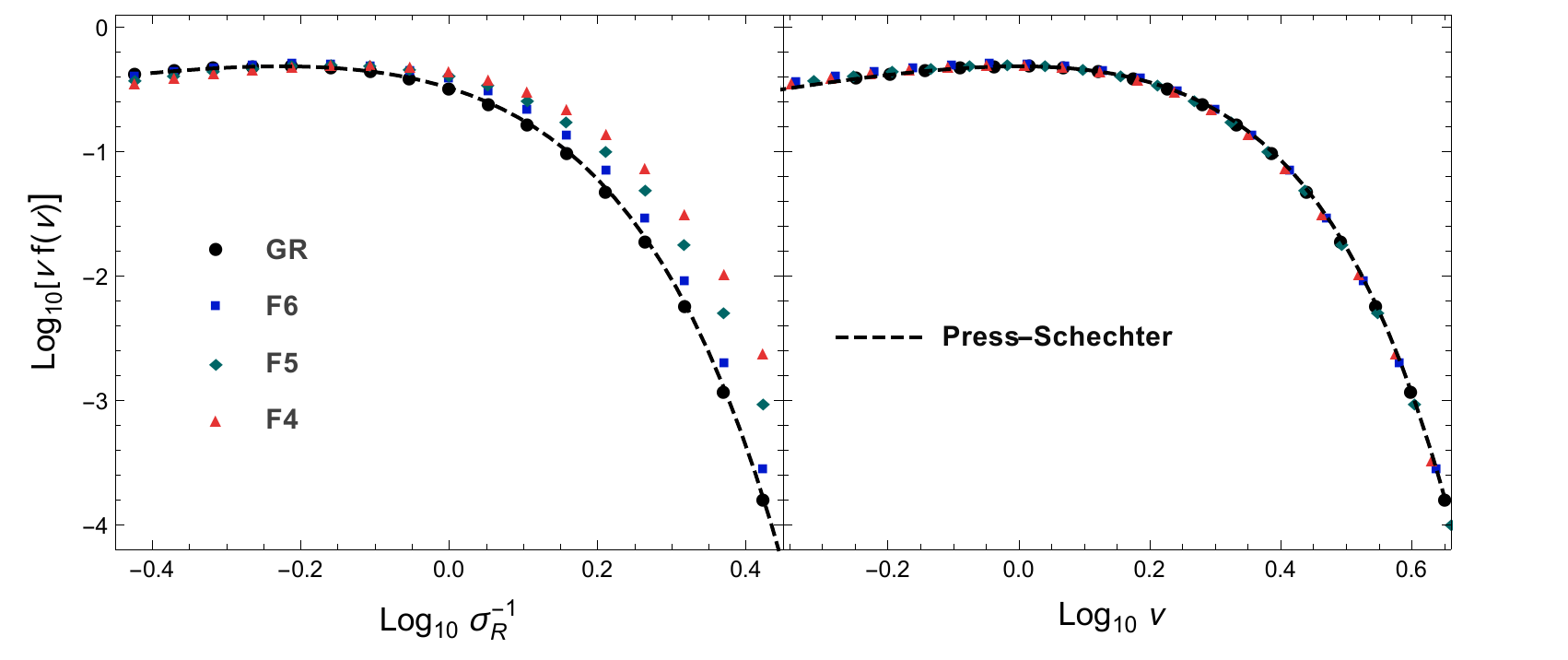}
	\caption{Multiplicity function $\nu f(\nu)$ as a function of $\log \sigma^{-1}_R$ (left panel), and as function of $\nu$ (right panel).
	We notice that while $\sigma_R$ is the same for all models, the peak significance $\nu$ is a model dependent quantity. The dashed curve
	shows the Press-Schechter analytical mass function. We have used the redshift for collapse $z_c=0$.
	\label{fig:massfunction}}
	\end{center}
\end{figure}

The different realizations of the overdensity produce an ensemble of
walkers which cross the barrier $\delta^\text{MG}_c(M,z)$ at different ``times'' $S$.  The main assumption in excursion set theory 
is that the halo mass function is directly related to the fraction 
$f_\text{FU}(S)dS$ of walkers that have their first (up-)crossing in the interval $(S,S+dS)$, 
\begin{equation}
 n(M,z) d M = \frac{\bar \rho}{M^2}  f_\text{FU}(S) dS, 
\end{equation}
that when comparing to eq.~(\ref{massfunction}) gives
\begin{equation}
f(\nu)d\nu = f_\text{FU}(S)dS \,,
\end{equation}
which can be used to compute the bias parameters (\ref{bnOfM}).

 \begin{figure}
	\begin{center}
	\includegraphics[width=3 in]{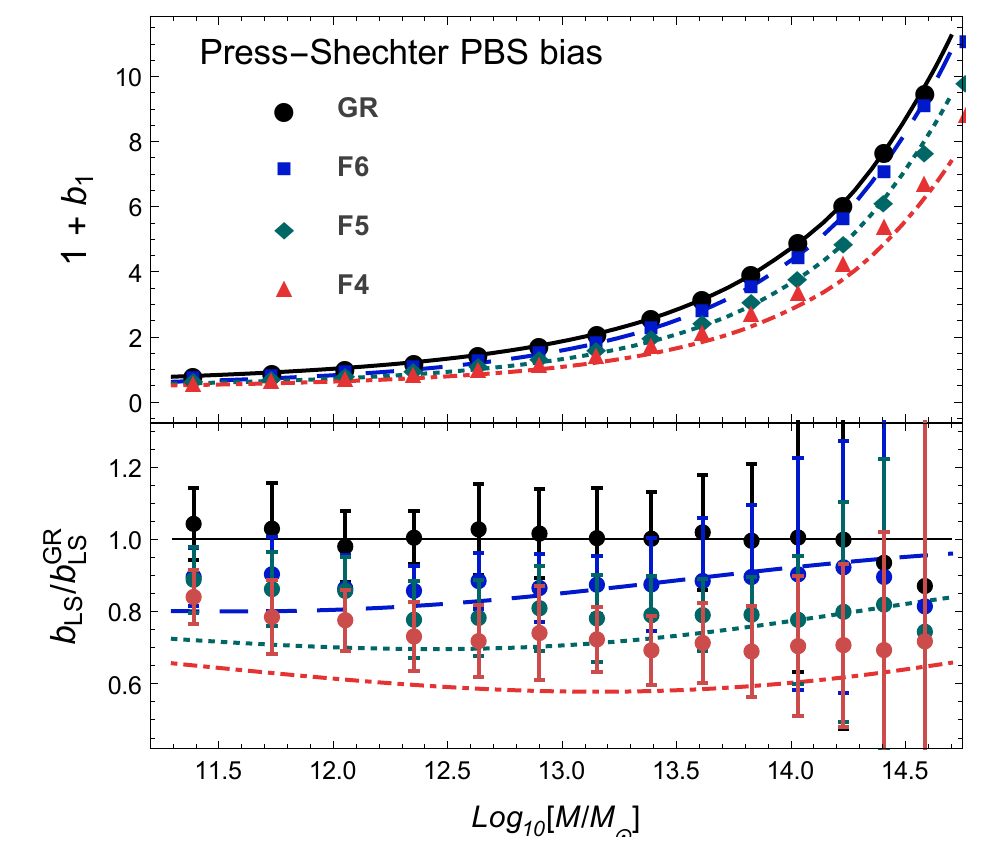}
	\caption{Large scales bias $b_\text{LS} =1+b_1$ for  models $\Lambda$CDM, F6, F5 and F4 as a function of halo mass $M$. The analytical results 
	make use of eq.~(\ref{b1}) with $p=0$ and $q=1$ corresponding to Press-Schechter PBS bias in GR, and use the threshold 
	density for collapse computed from the fitting functions provided in ref.~\cite{Kopp:2013lea}.  
	We plot solid black curves for $\Lambda$CDM model, dashed blue for F6, dotted green for F5, and 
	dotted-dashed for F4. The dots are the numerical results using excursion sets in section \ref{sec:ES}. The lower 
	paner shows the ratio of these quantities to the analytical $\Lambda$CDM bias. 
	\label{fig:b1ES}}
	\end{center}
\end{figure}

In order to estimate $f_\text{FU}$ we perform Monte Carlo simulations considering the simplest case of a sharp-$k$ filtering 
$\tilde{W}=\Theta_H(1/R(M) - k)$. 
In such a way the trajectories $\delta_R(S)$ are  Markovian, meaning that each jump
$\Delta \delta_R$ from $S$ to $S+\Delta S$ is uncorrelated with the previous ones, and are drawn from a Gaussian distribution. 
In $\Lambda$CDM this process can be solved analytically to obtain the Press-Schechter 
mass function and corresponding bias parameters. This is not the case for moving barriers, as those present in MG, and the solution should be found
numerically. 
We emphasize that we are considering this procedure at initial time with an already environmental independent density collapse, so our 
approach is that of  \cite{Kopp:2013lea}, but here we restrict our attention to the simplest 
case --- the authors of that work consider subsequent refinenemts
by accounting for top-hat window function, which has been discussed to be more proper \cite{Bond:1990iw}, as well as drifting and diffusing barriers 
to model non-spherical collapse.
Other approaches make this analysis for environmental dependent collapse densities and then average the resulting 
first up-crossing distributions  \cite{Li:2011qda,Li:2012ez,Lam2012,Lombriser:2013wta,vonBraun-Bates:2017usv}.

We plot our results in figure \ref{fig:massfunction}, showing the multiplicity function as a function of $\log \sigma^{-1}_R$ in the left
panel and as a function of $\nu$ in the right panel. Since we are taking the walkers at initial time, the variance, and hence $\sigma^{-1}_R$ is
equal to all models. But this is not the case for $\nu = \delta_c/\sigma$ which is model dependent. We note that for the latter case, all 
the multiplicity functions are (almost) equal, suggesting a universal pattern. When properly rescaled to  $\sigma^{-1}_R$, they look different. In
ref.~\cite{Kopp:2013lea} (their figure 9), the authors present similar results to our plot in the left panel of figure \ref{fig:massfunction}.

The local linear bias is computed by taking the logarithmic derivative of $\nu f(\nu)$ and dividing by the collapse threshold $\delta_c$. In 
figure \ref{fig:b1ES} the marks show our numerical results with error bars denoting the scattering after performing several runs
for walkers. Together, we plot the analytical results obtained  
from eq.~(\ref{b1}) using Press-Schechter parameters. For massive halos we find a decent agreement between our bias analytical formula and the numerics. For 
small masses our model underestimate the bias, though it follows correctly the trend of the data, as shown in the upper panel of figure \ref{fig:b1ES}.

\end{subsection}

\end{section}

\begin{section}{Conclusions}\label{sec:conclusions}

In this work we have continued the development of perturbation theory for MG theories within the Lagrangian formalism. Our main goal was to develop a theory of 
large-scale-structure bias for MG models that was lacking in the literature. However, to develop the theory it was first necessary  to understand the SPT power 
spectrum computed from the LPT formalism developed in a previous work \cite{Aviles:2017aor}.  In order to compute the SPT power spectrum,
we found relations that connect the Lagrangian and Eulerian kernels to arbitrary order in PT and that holds for general cosmological models,
these are given by
eqs.~(\ref{LPTtoSPTFn}) and (\ref{LPTtoGn}).  
Using these kernels, our result for the SPT power spectrum 
is given by eqs.~(\ref{RelP22RQ}, \ref{RelP13RQ}) that exactly coincides with the expression valid for the $\Lambda$CDM model. 
Of course, in eqs.~(\ref{RelP22RQ}, \ref{RelP13RQ}) one should employ the $P_L$, $Q$- and $R$-functions corresponding to the appropriate cosmology. 
This derivation was in fact not known, although 
it was perhaps expected and, in any case, necessary for us to develop bias contributions.    
In Ref.~\cite{Aviles:2017aor} two of us  have named the SPT power spectrum computed from eqs.~(\ref{RelP22RQ}, \ref{RelP13RQ}) as SPT*; now, we have 
demonstrated this result coincides with the SPT standard result. 

To finally develop the bias theory for MG cosmological models, we start from Lagrangian,  initially biased tracers.   
We consider smoothed overdensities of the 
underlying dark matter field and of the MG-related scalar field, and note that given the form of the Klein Gordon equation, 
the inclusion of bias expansion dependence on $\nabla^2 \varphi$ is degenerated with a bias dependence on 
$\nabla^2 \delta$. For that reason we assume a Lagrangian bias function with operators $\delta$ and $\nabla^2\delta$ as arguments
that will finally conduct us to define the bias 
parameters $b_{nm}$ and explain the way to renormalize them with the prescription given in \cite{Aviles:2018thp}.   
Then we computed the linear correlation function and power spectrum 
for biased tracers.
Next step is to construct the full 
LPT power spectrum and the CLPT correlation function 
for biased tracers in alternative cosmologies with screening mechanisms, in which we introduced local bias to 
second order and curvature bias to first order to match consistency in the 1-loop approximation order.  
By doing this we have generalized, beyond $\Lambda$CDM, the $Q$ and $R$ functions which are building blocks of 
matter and tracers statistics; in appendix D  we provide with all these functions. 

In order to facilitate the use of our formalism, that implies the manipulation of many cumbersome formulae, 
we are making public available a new code, called \verb|MGPT| (Modified Gravity Perturbation Theory),  that computes all 
the necessary functions and the SPT power spectrum and the CLPT correlation function for $\Lambda$CDM and MG models 
that can be brought to a scalar-tensor description, e.g. through field redefinitions; the code is released in the 
github repository \href{https://github.com/cosmoinin/MGPT}{cosmoinin/MGPT}.  To illustrate the functionality and tests of  code, 
we have computed these 2-point, 1-loop statistics for models $\Lambda$CDM, Hu-Sawicki $f(R)$ and DGP braneworld. 

Finally, we put forward a simple halo bias model to illustrate its effects stemming from modifications of gravity.    
We note that in general terms the MG dynamics are scale dependent and because the fifth force is attractive (in many models),  
matter fluctuations grow faster than in GR, leading to a more efficient relaxation of bias, and implying that
in general one can expect $ b_n^\text{MG}  <   b_n^\text{GR}$.
This result is generic and can 
contribute to distinguish MG models from $\Lambda$CDM.  To illustrate this fact, we constructed a 
particular bias model for the Hu-Sawicki $f(R)$ gravity and compare our analytical results with excursion set theory.  
Our results are consistent and shown in figure \ref{fig:b1ES}, confirming that $ b_n^{f(R)}  <   b_n^\text{GR}$ holds over a 
wide range of halo masses.

\end{section}

\appendix

\begin{section}{MG models}\label{app:mg}

This appendix aims to provide the required functions that are introduced in the formalism to determine the specific MG model. These functions are the so-called 
$M$ functions appearing in the Klein-Gordon equation and that are valid within the quasi-static limit, whose validity was 
studied elsewhere for both models considered here \cite{Song:2006ej, Khoury:2003rn, Sawicki:2006jj, Koyama:2005kd, Noller:2013wca}.  
The employed models are:  Hu-Sawicky $f(R)$ gravity and  DGP braneworld models.

\begin{subsection}{Hu-Sawicky $f(R)$ gravity model}\label{app:mg_HS}

This is the most studied $f(R)$ model, see details in ref.~\cite{Hu:2007nk}, in which the Einstein-Hilbert Lagrangian density is 
substituted by a general function $\sqrt{-g}(R + f(R))$ of the Ricci scalar.   
The model introduces a scalar degree of freedom, $\varphi= \delta f_R$, with $f_R=df/dR$, and is characterized by its value $f_{R0}=f_R|_{z=0}$,
that allow us to specify both a particular background cosmology 
and a fifth force range for the scalar, gravitational fifth force. Klein-Gordon equation for $\varphi$ can be found by taking the trace to the modified Einstein's field equations, an in the quasi-static limit this is
\begin{equation}
 \frac{3}{a^2}\nabla^2_\vx \varphi = -2 A_0  \delta + \delta R,
\end{equation}
with $\delta R = R(f_R) - R(\bar{f}_R)$. Comparing to eq.~(\ref{KGeq}) we note $\beta^2 = 1/6$, and $\mI = M_1 \varphi + \delta I = \delta R$. 
The $M$ functions are obtained by expanding $\delta R = M_1 \varphi +\frac{1}{2} M_2 \varphi^2 +\frac{1}{6} M_3 \varphi^3 +\cdots$, and by using
\begin{equation}
 R(f_{R})\simeq \bar{R} (f_{R0}/f_R)^{1/2},
\end{equation}
which is valid in cosmological scenarios, and where the background value of the Ricci scalar is $\bar{R}=3H_0^2(\Omega_{m0}a^{-3} + 4 \Omega_\Lambda)$. We obtain
\begin{align}\label{M1fR}
M_1(a) = \frac{3}{2}  \frac{H_0^2}{|f_{R0}|} \frac{(\Omega_{m0} a^{-3} + 4 \Omega_\Lambda)^3}{(\Omega_{m0}  + 4 \Omega_\Lambda)^2}, 
\end{align}
\begin{align}\label{M2fR}
M_2(a) = \frac{9}{4}  \frac{H_0^2}{|f_{R0}|^2} \frac{(\Omega_{m0} a^{-3} + 4 \Omega_\Lambda)^5}{(\Omega_{m0}  + 4 \Omega_\Lambda)^4}, 
\end{align}
\begin{align}\label{M3fR}
M_3(a)=  \frac{45}{8}  \frac{H_0^2}{|f_{R0}|^3} \frac{(\Omega_{m0} a^{-3} + 4 \Omega_\Lambda)^7}{(\Omega_{m0}  + 4 \Omega_\Lambda)^6}, 
\end{align}
while the mass is given by $m=\sqrt{M_1/3}$.
Values $f_{R0}=-10^{-4}, -10^{-5}, -10^{-6}$ are used in this paper and correspond to models
F4, F5 and F6, respectively. The fact that for these values of $f_{R0}$ the expansion history is indistinguishable to that in $\Lambda$CDM, as we have assumed, has been studied in \cite{Hu:2007nk}.

\end{subsection}

\begin{subsection}{DGP braneworld model}\label{app:mg_DGP}

The DGP model proposes we are living 4-D brane immersed in a 5-D spacetime \cite{Dvali:2000hr} 
in which the interesting parameter is the crossover scale ($r_c$) that is proportional to the ratio of the  5-D to 4-D gravitational constants.
The Hubble flow is given by
\begin{equation}
 H(z) = H_0 \left( \sqrt{\Omega_{m0}(1+z)^3 + \Omega_r} + \epsilon \sqrt{ \Omega_r}  \right),
\end{equation}
with $\Omega_r = 1/4r_c^2 H_0^2$. It has two branches: the self-accelerating (sDGP) \cite{Deffayet:2000uy}, corresponding to $\epsilon=1$,  
and the normal branch (nDGP), corresponding to $\epsilon=-1$.
Both models have interesting features that have been intensively studied, see 
e.g. \cite{Song:2007wd}.  We choose in this work the nDGP model, though unfortunately  its 
background evolution is very different than in $\Lambda$CDM. However, one can add a smooth
dark energy component to match as much as desired the background evolution in $\Lambda$CDM \cite{Schmidt:2009sv}. 

This model possesses no mass term, hence $M_1=0$ and $m=0$, but it has  $k$ dependences for higher order terms stemming 
from quadratic second derivatives in the Klein-Gordon equation: 
\begin{align} \label{KGDGP}
 \frac{1}{a^2}\nabla^2_\vx\varphi &= -  4 A_0 \beta^2 \delta  +  2\beta^2 \frac{r^2_c}{a^4}\Big( (\nabla^2_x \varphi)^2 - (\nabla_{\vx\,i} \nabla_{\vx\,j}\varphi)^2\Big) ,
\end{align}
hence, comparing to eq.~(\ref{KGeq}),
\begin{align}
 \delta \mI &= 
  \frac{r^2_c}{a^4}\Big[ (\nabla^2_x \varphi)^2 - (\nabla_{\vx\,i} \nabla_{\vx\,j}\varphi)^2\Big] \nonumber\\  
 &= \frac{r^2_c}{a^4}\Big[ ( \varphi_{,ii})^2 - (\varphi_{,ij})^2 
    - 4 \Psi_{i,m} \varphi_{,im} \varphi_{,jj} - 2 \Psi_{i,mi} \varphi_{,m} \varphi_{,jj} \nonumber\\
 & \qquad   + 4 \Psi_{i,m} \varphi_{,mj} \varphi_{,ij} + 2 \Psi_{j,mi} \varphi_{,m} \varphi_{,ij}\Big], 
\end{align}
where the second equality arises when transforming from Eulerian to Lagrangian coordinates. The Fourier transform of Lagrangian displacements give
$[\Psi_{i,j}](\vk) = -\frac{k_i k_j}{4 a^2 A_0 \beta^2}\varphi(\vk)$ to leading order. Comparing to eq.~(\ref{deltaI}) we can read the $M$ functions as
\begin{align}
 M_2(\vk_1,\vk_2) &= \frac{2 r_c^2}{a^4} \left[ k_1^2 k_2^2 - (\vk_1 \cdot \vk_2)^2 \right], \\
 M_3(\vk_1,\vk_2,\vk_3) &= \frac{3 r_c^2}{a^6 A_0 \beta^2} \Big( 
   2 (\vk_1 \cdot \vk_2)^2 k_3^2 + (\vk_1 \cdot \vk_2 ) k_1^2 k_3^2  \nonumber\\
  &\qquad - (\vk_1 \cdot \vk_2)(\vk_1\cdot\vk_3)^2 - 2(\vk_1\cdot\vk_2)(\vk_2\cdot\vk_3)(\vk_3\cdot\vk_1) \Big). 
\end{align}
%
In addition, we have $A(t) =  A_0 \, \left( 1+ 2 \beta^2 \right)$,
with
\begin{align}\label{betaDGP}
\beta^2(t) &= \frac{1}{6 \beta_\text{DGP}(t)}, \quad
\beta_\text{DGP}(t) = 1 - 2 \epsilon H r_c \left(1+\frac{\dot{H}}{3 H^2} \right).
\end{align}
In \text{DGP} literature a different notation for $\beta$ is commonly used, in which $[\beta_\text{DGP}]^\text{here}=[\beta]^\text{other works}$. 

\end{subsection}

\end{section}

\begin{section}{$G_{n}$ kernels}\label{app:Gn}
 
Computing the $G_n$ kernels is a little more messy than the $F_n$, but still straightforward.
The velocity field is given by $v_i(\vx,t) = d\vx/dt = \dot{\Psi}_i(\vq,t)$.
We use $\partial/\partial q^i = (\partial x^j /\partial q^i)\partial/ \partial x^j = J_{ji}\partial/ \partial x^j $,
to get $\nabla_{\vx i} = (J^{-1})_{ji} \nabla_{ j}$, with 
$(J^{-1})_{ji} = \epsilon_{ikp}\epsilon_{jqr}J_{kq}J_{pr}/ 2 J $ or
\begin{equation}
 J(J^{-1})_{ji} =  \delta_{ij} + (\delta_{ij}\delta_{ab} - \delta_{ia}\delta_{jb})\Psi_{a,b} + \frac{1}{2}\epsilon_{ikp}\epsilon_{jqr} \Psi_{k,q} \Psi_{p,r} ,
\end{equation}
hence
 \begin{equation}
 J \nabla_{\vx\,i}v_i = \dot{\Psi}_{i,i} + \Psi_{j,j}\dot{\Psi}_{i,i} - \Psi_{i,j}\dot{\Psi}_{i,j} 
 + \frac{1}{2}\epsilon_{ikp}\epsilon_{jqr} \Psi_{k,q} \Psi_{p,r}\dot{\Psi}_{i,j}.
 \end{equation}
The Fourier transform of the velocity divergence yields
\begin{align}
 H \theta(\vk) &= \int d^3 x e^{-i\vk \cdot \vx}  \nabla_{\vx\,i}v_i 
               =  \int d^3 q e^{-i\vk \cdot \vq} e^{-i\vk\cdot \Psi(\vq,t)} J(\vq,t)  \nabla_{\vx\,i}v_i \nonumber\\
              &=  \int d^3 q e^{-i\vk \cdot \vq} \sum_{\ell=0}^{\infty} (-ik_a \Psi_a)^{\ell}  \dot{\Psi}_{i,i}  
              +  (\delta_{ij}\delta_{ab} - \delta_{ia}\delta_{jb}) \int d^3 q e^{-i\vk \cdot \vq} \Psi_{a,b}\dot{\Psi}_{i,j} \sum_{\ell=0}^{\infty} (-ik_k \Psi_k)^{\ell}   \nonumber\\
              &+  \frac{1}{2}\epsilon_{ikp}\epsilon_{jqr}  \int d^3 q e^{-i\vk \cdot \vq}   \Psi_{k,q} \Psi_{p,r}\dot{\Psi}_{i,j} \sum_{\ell=0}^{\infty} (-ik_s \Psi_s)^{\ell}.  
\end{align}

On the other hand by taking the derivative of the displacement field in eq.~(\ref{PsiExp})
\begin{align}
 \dot{\Psi}_i(\vp) &=   i \sum_{m=1}^{\infty} \frac{1}{m!} \underset{\vp_{1\cdots m} = \vp}{\int}  L_{i}^{'(m)}(\vp_1,\dots,\vp_m)
 \delta_L(\vp_1) \cdots \delta_L(\vp_{n}), 
\end{align}
with
\begin{equation}
 L_{i}^{'(m)}(\vp_1,\dots,\vp_m) =  \dot{L}_{i}^{(m)}(\vp_1,\dots,\vp_m) + H L_{i}^{(m)}(\vp_1,\dots,\vp_m)(f(\vp_1) + \cdots + f(\vp_m)).
\end{equation}
We get
\begin{flalign} \label{LPTtoGn}
H G_n(\vp_1,\dots\vp_n) =  \sum_{\ell=0}^{n-1} \sum_{m_1+\dots+ m_{\ell+1} = n }  
\frac{k_{i_1}\cdots k_{i_\ell} a_i}{\ell! m_1!\cdots m_{\ell+1}! }  L^{(m_1)}_{i_1}(\vp_1,\dots,\vp_{m_1}) \cdots   
&&\nonumber\\ 
  \cdots L^{(m_\ell)}_{i_\ell} (\vp_{m_{\ell-1} +1},\dots,\vp_{m_\ell}) 
    L^{'(m_{\ell+1})}_{i} (\vp_{m_{\ell }+1},\dots,\vp_{m_{\ell+1}}) 
&&\nonumber\\
      - (\delta_{ij}\delta_{pq} - \delta_{ip}\delta_{jq}) \sum_{\ell=0}^{n-2} \sum_{m_1+\dots+ m_{\ell+2} = n }   
  \frac{k_{i_1}\cdots k_{i_\ell} a_q b_j}{\ell! m_1!\cdots m_{\ell+2}! } L^{(m_1)}_{i_1}(\vp_1,\dots,\vp_{m_1}) \cdots 
&&\nonumber\\     
 \cdots L^{(m_\ell)}_{i_\ell} (\vp_{m_{\ell-1} +1},\dots,\vp_{m_\ell})  L^{(m_{\ell+1})}_{p} (\vp_{m_{\ell }+1},\dots,\vp_{m_{\ell+1}}) 
  L^{'(m_{\ell+2})}_{i} (\vp_{m_{\ell+1 }+1},\dots,\vp_{m_{\ell+2}})   
&&\nonumber\\
  + \frac{1}{2}\epsilon_{ikp}\epsilon_{jqr}  \sum_{\ell=0}^{n-3} \sum_{m_1+\dots+ m_{\ell+3} = n }  
  \frac{k_{i_1}\cdots k_{i_\ell} a_r b_q c_j}{\ell! m_1!\cdots m_{\ell+3}! } L^{(m_1)}_{i_1}(\vp_1,\dots,\vp_{m_1}) \cdots  
L^{(m_\ell)}_{i_\ell} (\vp_{m_{\ell-1} +1},\dots,\vp_{m_\ell}) 
&& \nonumber\\ 
  \times L^{(m_{\ell+1})}_{p} (\vp_{m_{\ell }+1},\dots,\vp_{m_{\ell+1}}) 
  L^{(m_{\ell+2})}_{k} (\vp_{m_{\ell }+2},\dots,\vp_{m_{\ell+2}}) L^{'(m_{\ell+3})}_{i} (\vp_{m_{\ell+2 }+1},\dots,\vp_{m_{\ell+3}}), &&
\end{flalign}
where $\vk$ is the sum of all involved momenta, $\vk = \vp_1 + \cdots + \vp_n$ for order $n$ kernel, and
\begin{align}
 a_{i} &= (\vp_{m_{\ell} +1}+\cdots + \vp_{m_{\ell+1}})_i, \\
 b_{i} &=(\vp_{m_{\ell+1} +1}+\cdots + \vp_{m_{\ell+2}})_i, \\
 c_{i} &=(\vp_{m_{\ell+2} +1}+\cdots + \vp_{m_{\ell+3}})_i.  
\end{align}
The notation may be somewhat confusing, for LPT kernels that are contracted with momentum $\vk$ (these are the corresponding to 
$m_1$ to $m_{\ell}$) we allow $m_i$ to take zero values, in such a case we define $L^{(m_i=0)}_{i_l} \equiv 1$. While kernels contracted with momenta $\ve a$, 
 $\ve b$ and  $\ve c$  are restricted to have positive orders, $m_{\ell} + 1,\dots, m_{\ell+3} \geq1$.

\end{section}

\begin{section}{$k$- and $q$-functions}\label{app:kqfunctions}

In this appendix we give expressions for the $q$- and $k$-functions of Sects.\ref{sec:2} and \ref{sec:3}. 
Before we proceed to display all of them we show how 
they arise by considering $A_{ij}$ as an example: 
\begin{align}
A_{ij}&= \langle \Delta_i\Delta_j\rangle_c = \int \Dk{k_1}\Dk{k_2} (e^{i\vk_1\cdot\vq_2}-e^{i\vk_1\cdot\vq_1})
 (e^{i\vk_2\cdot\vq_2}-e^{i\vk_2\cdot\vq_1}) \langle \Psi_i(\vk_1) \Psi_j(\vk_2) \rangle_c.
\end{align}
By homogeneity and rotational symmetry we have
\begin{equation}
 \langle \Psi_i(\vk) \Psi_j(\vk') \rangle_c =  (2\pi)^3 \delta_\text{D} (\vk + \vk') \left(\delta_{ij} p(k) + \frac{k_i k_j}{k^2} a(k) \right)
 = (2\pi)^3 \delta_\text{D} (\vk + \vk')\frac{k_i k_j}{k^2} a(k), 
\end{equation}
where in the last equality we use our assumption that the Lagrangian displacement are longitudinal, $\Ps = \vk (\vk \cdot \Ps)$.
By expanding perturbatively $\langle \Psi_i(\vk_1) \Psi_j(\vk_2) \rangle_c$ and using the definitions of eqs.~(\ref{R1}, \ref{Q1}) below we have 
\begin{equation}\label{Aij1}
 A_{ij}(\vq) = \int \Dk{k} (2 - e^{i \vk \cdot \vq}- e^{-i \vk \cdot \vq}) \frac{k_ik_j}{k^4} \left(P_L(k) + \frac{3}{7}Q_1(k) + \frac{10}{21} R_1(k)\right).
\end{equation}
In analogous way all $q$-functions can be obtained.

\begin{subsection}{$k$-functions}\label{app:kfunctions}

Let us define the mixed polyspectra at order $m + n_1 +\cdots n_N$
\begin{align}
\langle \delta_L(\vk_1)\cdots\delta_L(\vk_m) \Psi_{i_1}^{(n_1)}(\vp_1)\cdots\Psi_{i_N}^{(n_N)}(\vp_N)\rangle_c &= \nonumber \\
- (-i)^N C_{i_1\cdots i_N}^{(n_1\dots n_N)}(\vk_1,\cdots\vk_{m},;\vp_1,\cdots,\vp_N) 
& (2 \pi)^3 \delta_\text{D}(\vk_1+\cdots+\vp_N),
\end{align}
which is composed of $m$ smoothed linear overdensities and $N$ Lagrangian displacements.
This definition differs by a minus sign to that in \cite{Matsubara:2008wx}, 
but coincides with \cite{Matsubara:2007wj} if bias is not considered, that is for $m=0$.
The following polyspectra are needed
\begin{align}
 C_{ij}^{(11)}(\vk) &= L^{(1)}_i(\vk) L^{(1)}_j(\vk)P_L(k),  \\
 C_{ij}^{(22)}(\vk) &=\frac{1}{2} \int \Dk{p} k_i k_j L_i^{(2)}(\vk - \vp,\vp) L_i^{(2)}(\vk - \vp,\vp) P_L(|\vk-\vp|)P_L(p), \label{Cij22} \\
 C_{ij}^{(13)}(\vk) &= \frac{1}{2} L^{(1)}_i(\vk) P_L(k) \int \Dk{p} L^{(3)}_j(\vk,-\vp,\vp)P_L(p),\\
 C_{ijk}^{(112)}(\vk_1,\vk_2,\vk_3) &=C_{jki}^{(121)}(\vk_2,\vk_3,\vk_1) = C_{kij}^{(211)}(\vk_3,\vk_1,\vk_2) \nonumber\\ 
           & = -L^{(1)}_i(\vk_1)L^{(1)}_j(\vk_2)L^{(2)}_k(\vk_1,\vk_2)P_L(k_1)P_L(k_2), \label{Cij13} \\
 C_{ij}^{(12)}(\vp_1;\vp_2,\vp_3) &= C_{ji}^{(21)}(\vp_1;\vp_3,\vp_2) = 
 L^{(1)}_i(\vp_2) L_j^{(2)}(\vp_1,\vp_2)P_L(p_1)P_L(p_2), \label{Cij12}\\
 C^{(2)}_i(\vp_1,\vp_2;\vp_3) &= - L^{(2)}_i(\vp_1.\vp_2) P_L(p_1) P_L(p_2). \label{Cij2}\\
\end{align}

The only $k$-function that involves third order fluctuation is
\begin{align}
 R_1(k) &\equiv  \frac{21}{5} k_i k_j  C_{ij}^{(13)}(\vk)  
 = \int \Dk{p} \frac{21}{10} k_i L^{(3)s}_{i}(\vk,-\vp,\vp) P_L(p)P_L(k) \label{R1}\\
  &=\int \Dk{p} \frac{21}{10} \frac{D^{(3)s}(\vk,-\vp,\vp)}{D_{+}(k)D_{+}^2(p)} P_L(p)P_L(k).
\end{align}
Label ``s'' means that $D^{(3)}$ should be symmetrized over; its expression is somewhat large and we do not present it here, 
it is given by eq.~(84) of \cite{Aviles:2017aor}. 

We define the ratio of internal over external momenta and the cosine of the angle between them as $r=p/k$ and $x= \hat{\vk} \cdot \hat{\vp}$. 
The $k$-functions constructed with both linear and second order displacement fields are 
\begin{align}
  Q_{1}(k)   &\equiv \frac{98}{9} k_i k_j C_{ij}^{(22)}(\vk) \\
 &=  \int \Dk{p} \left(\A - \B\frac{(\vp\cdot(\vk-\vp))^2}{p^2|\vk -\vp|^2} \right)^2  P_L(|\vk-\vp|)P_L(p) \\
 &= \frac{k^3}{4 \pi^2}\int_0^\infty dr P_L(kr)\int_{-1}^1 dx r^2 
            \left(\A - \B\frac{(-r+x)^2}{1+r^2-2rx} \right)^2 P_L(k\sqrt{1+r^2-2rx}), \label{Q1}
\end{align}
\begin{flalign}
 Q_{2}(k)  &\equiv \frac{7}{3}k_i k_j k_k \int \Dk{p} C_{ijk}^{(211)}(\vk,-\vp,\vp-\vk)  &&\\ 
 &=  \int \Dk{p} \frac{\vk \cdot \vp \, \,\vk \cdot (\vk-\vp)}{p^2 |\vk-\vp|^2}
      \left(\A - \B\frac{(\vp\cdot(\vk-\vp))^2}{p^2|\vk -\vp|^2} \right)  P_L(|\vk-\vp|)P_L(\vp) &&\\
 &= \frac{k^3}{4\pi^2} \int_0^\infty dr P_L(kr) \int_{-1}^1 d x \frac{rx(1-rx)}{1+r^2-2rx}
          \left(\A-\B\frac{(r-x)^2}{1+r^2-2rx}\right)    P_L(k\sqrt{1+r^2-2rx}), && \label{Q2}
\end{flalign}
\begin{flalign}
 Q_{I}(k)   &\equiv \frac{7}{3} (k_i k_j k_k - k^2 k_i \delta_{jk}) \int \Dk{p} C_{ijk}^{(211)}(\vk,-\vp,\vp-\vk) &&\nonumber\\
            &= \int \Dk{p} \frac{(\vk \cdot \vp)\vk \cdot (\vk-\vp)- k^2 \vp \cdot (\vk-\vp)}{p^2 |\vk-\vp|^2}
              \left(\A - \B\frac{(\vp\cdot(\vk-\vp))^2}{p^2|\vk -\vp|^2} \right)  P_L(|\vk-\vp|)P_L(p)  &&\nonumber\\
            &= \frac{k^3}{4 \pi^2}\int_0^\infty dr P_L(kr)\int_{-1}^1 dx \frac{r^2(1-x^2)}{1+r^2 - 2 r x} 
            \left(\A - \B\frac{(-r+x)^2}{1+r^2-2rx} \right) P_L(k\sqrt{1+r^2-2rx}),  &&\label{QI}
\end{flalign}
\begin{flalign}
 Q_5(k)     &\equiv \frac{7}{3} k_ik_j \int \Dk{p}    C^{(12)}_{ij}(-\vp;\vp- \vk,\vk) &&\nonumber\\
 &=\int \Dk{p} \frac{\vk \cdot \vp}{p^2}  \left(\A - \B\frac{(\vp\cdot(\vk-\vp))^2}{p^2|\vk -\vp|^2} \right) P_L(p)P_L(|\vk- \vp|) &&\nonumber\\
       &=\frac{k^3}{4\pi^2} \int_0^{\infty}  P_L(kr) \int_{-1}^1 dx 
          r x \left(\A - \B\frac{(-r+x)^2}{1+r^2-2rx} \right) P_L(k\sqrt{1+r^2-2rx}), &&\label{Q5} 
\end{flalign}
\begin{flalign}
 Q_8(k) &\equiv \frac{7}{3} k_i \int \Dk{p} C_i^{(2)}(-\vp,\vp - \vk;\vk)   &&\nonumber\\
        &= \int \Dk{p} \left(\A - \B\frac{(\vp\cdot(\vp-\vk))^2}{p^2|\vp -\vk|^2} \right) P_L(p)P_L(|\vp- \vk|) &&\nonumber\\
        &= \frac{k^3}{4\pi^2}\int_0^{\infty} dr P_L(kr) \int_{-1}^1 dx r^2\left(\A-\B\frac{(r-x)^2}{1+r^2-2rx}\right)  P_L(k\sqrt{1+r^2-2rx}), &&\label{Q8}
\end{flalign}

\begin{flalign}
 R_{2}(k)   &\equiv \frac{7}{3}k_i k_j k_k \int \Dk{p} C_{ijk}^{(112)}(\vk,-\vp,\vp-\vk)  &&\nonumber\\ 
 &=     \int \Dk{p} \frac{\vk \cdot \vp \, \,\vk \cdot (\vk-\vp)}{p^2 |\vk-\vp|^2}\left(\A - \B\frac{(\vp\cdot \vk)^2}{p^2k^2} \right)  P_L(k)P_L(p)  && \nonumber \\
 &=    \frac{k^3}{4\pi^2} P_L(k) \int_0^\infty dr P_L(kr) \int_{-1}^1 \frac{rx(1-rx)}{1+r^2 - 2rx}\left(\A - \B x^2 \right) , &&\label{R2}
\end{flalign}
\begin{flalign}
 R_{I}(k)   &\equiv \frac{7}{3}(k^2\delta_{ij} - k_ik_j) \int \Dk{p} C_{ij}^{(12)}(\vk;-\vp,\vp-\vk) &&\nonumber\\
 &= \int \Dk{p} \left[ \frac{\big( (\vk\cdot\vp) \vk -  k^2 \vp \big) \cdot(\vk-\vp) }{p^2|\vk-\vp|^2}\right] 
              \left(\A - \B\frac{(\vk\cdot\vp)^2}{p^2k^2} \right) P_L(p)P_L(k) &&\nonumber\\
 &= \frac{k^3}{4 \pi^2}P_L(k) \int_0^{\infty} dr P_L(kr) \int_{-1}^{1} dx \frac{r^2 (1-x^2)}{1+r^2-2 r x} \left( \A - \B x^2\right), && \label{RI}
\end{flalign}
\begin{flalign}
 R_{1+2}(k) &\equiv\frac{7}{3} k_i k_j \int \Dk{p}   C^{(12)}_{ij}(-\vp;\vk,\vp-\vk)  &&\nonumber\\
 & = \int \Dk{p} \frac{\vk \cdot (\vk - \vp)}{|\vk-\vp|^2}  \left(\A - \B\frac{(\vk\cdot\vp)^2}{p^2k^2} \right) P_L(p)P_L(k)  &&\nonumber\\
&=   \frac{k^3}{4\pi^2} \int_0^\infty dr\,  P_L(kr) \int_{-1}^{1}dx 
     \frac{r^2(1 - x r)}{1+r^2-2rx}(\A-\B x^2). &&
\end{flalign}
The normalized second order growth functions are evaluated as $\A,\B(\vp,\vk-\vp)$ 
for $Q$ functions,  while as $\A,\B(\vk,-\vp)$ for $R$ functions. For EdS (or more precisely for $\Lambda$CDM with $f=\Omega_m^{1/2}$), 
$\A=\B=1$ and we have the following relations
\begin{equation}
 R_I^\text{EdS} = R_1^\text{EdS},\quad  Q_I^\text{EdS} = Q_1^\text{EdS},\quad  R_{1+2}^\text{EdS} = R_1^\text{EdS} + R_{2}^\text{EdS}
\end{equation}
recovering the results of \cite{Matsubara:2007wj,Matsubara:2008wx,Carlson:2012bu}.

We further have $k$-functions constructed from linear displacements fields, hence these functions have the same form in $\Lambda$CDM and in MG,
\begin{align}
 Q_{3}(k) &=  \int \Dk{p} \frac{(\vk \cdot \vp)^2 (\vk\cdot (\vk-\vp))^2}{p^4 |\vk-\vp|^4} P_L(p)P_L(|\vk-\vp|),\\
 Q_7(k)    &= \int \Dk{p} \frac{(\vk \cdot \vp)^2}{p^4} \frac{\vk \cdot (\vk -\vp)}{|\vk -\vp|^2} P_L(p) P_L(|\vk -\vp|),\\
 Q_{9}(k)  &=  \int \Dk{p}    \frac{\vk \cdot \vp}{p^2} \frac{\vk \cdot (\vk - \vp)}{|\vk - \vp|^2}  P_L(p) P_L(|\vk - \vp|),\\
 Q_{11}(k) &= \int \Dk{p} \frac{(\vk \cdot \vp)^2}{p^4}  P_L(p) P_L(|\vk -\vp|), \\
 Q_{12}(k) &= \int \Dk{p}    \frac{\vk \cdot \vp}{p^2}  P_L(p) P_L(|\vk - \vp|), \\
 Q_{13}(k) &=  \int \Dk{p}  P_L(p) (P_L(|\vk -\vp|) - P_L(p)).
\end{align}
A direct computation leads to 
$Q_{13}(k) = \mathcal{F}[\xi_L^2(\vq)] = \int \Dk{p}  P_L(p) P_L(|\vk -\vp|)$ (see eq.(\ref{Q13der})). The 
constant term is added in order to make $Q_{13}$ insensitive to the smoothing scale \cite{Aviles:2018thp}, 
following the renormalization method of \cite{McDonald:2006mx}.

\end{subsection}

\begin{subsection}{$q$-functions}\label{app:qfunctions}

We first consider those $q$-functions that are required for the unbiased case. 
$A_{ij} \equiv \langle\Delta_i\Delta_j\rangle_c$ was already given by
eq.~(\ref{Aij1}), but here we exploit rotational symmetry to write it more conveniently as
one dimensional integrals,
\begin{equation}\label{Aij2}
 A_{ij}(\vq) = X(q) \delta_{ij} + Y(q) \hat{q}_i \hat{q_j},
\end{equation}
with
\begin{align}
X(q) &=  \frac{1}{\pi^2}\int_0^\infty dk \left(P_L(k) +\frac{9}{98}Q_1(k) + \frac{10}{21} R_1(k) \right)\left(\frac{1}{3} - \frac{j_1(kq)}{kq} \right), \\
Y(q) &=  \frac{1}{\pi^2}\int_0^\infty dk \left(P_L(k) +\frac{9}{98}Q_1(k) + \frac{10}{21} R_1(k) \right) j_2(kq),
\end{align}
where we used the definition of the polyspectra and the $k$-functions of the previous subsection.
Similarly for $W_{ijk} \equiv \langle\Delta_i\Delta_j\Delta_k\rangle_c$, see \cite{Vlah:2015sea},
\begin{align}\label{wijk}
W_{ijk}(\vq) = V(q)\hat{q}_{\{i}\delta_{jk\}} + T(q)\hat{q}_i\hat{q}_j\hat{q}_k,
\end{align}
with
\begin{align}
 V(q) &= -\frac{1}{\pi^2}\int_{0}^{\infty} \frac{dk}{k} \tilde{V}(k) j_1(kq)   - \frac{1}{5} T(q),  \label{wijkV} \\
 T(q) &= -\frac{1}{\pi^2}\int_{0}^{\infty} \frac{dk}{k}\tilde{T}(k) j_3(kq), \label{wijkT}
\end{align}
and
\begin{align}
 \tilde{V}(k) &= \frac{3}{35}(Q_I(k) - 3 Q_2(k) + 2 R_I(k) - 6 R_2(k)  ), \label{Vtilde}\\
 \tilde{T}(k) &= \frac{9}{14}(Q_I(k) + 2 Q_2(k) + 2 R_I(k) + 4 R_2(k)  ). \label{Ttilde}
\end{align}

Now we focus our attention to those $q$-functions that are accompanied by bias parameters. The most  
cumbersome of these is $A_{ij}^{10} = A_{ij}^{10(12)} + A_{ij}^{10(21)} = 2 A_{ij}^{10(12)} 
= 2 \langle \delta_L(\vq_1) \Delta_{i}^{(1)}\Delta_{i}^{(2)} \rangle$, or  
\begin{align}\label{Aij101}
 A_{ij}^{10} &= 2 \int \Dk{k_1} \Dk{k_2} \Dk{k_3} e^{i\vk_1 \cdot \vq_1} 
(e^{i\vk_2 \cdot \vq_2} - e^{i\vk_2 \cdot \vq_1})(e^{i\vk_3 \cdot \vq_2} - e^{i\vk_3 \cdot \vq_1})
  \langle \delta(\vk_1) \Psi^{(1)}_i(\vk_2)\Psi^{(2)}_j(\vk_3)\rangle_c \nonumber\\
 &= 2 \int \Dk{k} e^{i\vk\cdot\vq} \int \Dk{p} \Big[ (1+e^{-i\vk\cdot\vq} )  C_{ij}(-\vk;\vk-\vp,\vp)
    -  C_{ij}(-\vp;\vk,\vp-\vk) \nonumber\\
 &\qquad   -  C_{ij}(-\vp;\vp-\vk,\vk)\Big] \nonumber\\
 &= \int \Dk{k} e^{i\vk\cdot\vq} \frac{3}{14} \left[\frac{\delta_{ij}}{k^2}R_I(k)  -\frac{k_ik_j}{k^4}(2 R_{1+2}(k) + R_I(k) + 2 R_2(k) + 2 Q_5(k)) \right] \nonumber\\
       &\qquad + \int \Dk{k}  \frac{3}{14} \left[\frac{\delta_{ij}}{k^2}R_I(k)  -\frac{k_ik_j}{k^4}(R_I(k) + 2 R_2(k) ) \right],
\end{align}
where we have used the definition of $Q(k)$ and $R(k)$ functions in appendix \ref{app:kfunctions}. 
We can rewrite eq.~(\ref{Aij101}) in a simpler form
\begin{equation}
 A_{ij}^{10}(\vq) = X_{10}(q)\delta_{ij} + Y_{10}(q)\hat{q}^i \hat{q}^j,
\end{equation}
with
\begin{align}
X_{10}(q) &= \frac{1}{14 \pi^2}\int_0^\infty dk \left[ 2  (R_I-R_2) + 3   R_I j_0(kq) 
- 3( R_I + 2 R_2 + 2 R_{1+2} + 2 Q_5) \frac{j_1(kq)}{kq} \right], \\
Y_{10}(q) &= \frac{3}{14 \pi^2}\int_0^\infty dk  ( R_I +2 R_2 + 2 R_{1+2} + 2 Q_5 ) j_2(kq).
\end{align}

Analogous computations for functions $U_i^{mn}(\vq) = \langle \delta_L^m(\vq_1)\delta_L^n(\vq_2) \Delta_i \rangle_c  \equiv U^{mn}(q) \hat{q}_i$ give
\begin{align}
 U(q)       &=  - \frac{1}{2 \pi^2}\int_0^{\infty} dk \,k \left( P_L(k) + \frac{5}{21}R_1(k) \right) j_1(kq),    \\
 U^{20}(q)  &= -\frac{3}{14 \pi^2}\int_0^{\infty} dk \,k  Q_8(k) j_1(kq), \\
 U^{11}(q)  &= -\frac{3}{7 \pi^2}\int_0^{\infty} dk \,k  R_{1+2}(k) j_1(kq).
\end{align}

\end{subsection}

\end{section}

\begin{section}{SPT power spectrum}\label{app:PS}

Expanding the exponential $\exp (-\frac{1}{2}k_ik_j A_{ij} - \frac{i}{6} k_ik_jk_k W_{ijk} )$ in the tracers LPT power spectrum of eq.~(\ref{XLPTPS}) we obtain 
\begin{align}
 P^\text{SPT}_X(k) &= \int d^3 q e^{i \vk \cdot \vq } \Bigg[-\frac{1}{2}k_ik_j A_{ij} - \frac{i}{6} k_ik_jk_k W_{ijk} +\frac{1}{8}k_ik_jk_k k_l A^L_{ij}A^L_{kl}
    + b_{10}^2 \xi_L  + 2 i b_{10} k_i U_i
 \nonumber\\
 & \qquad   -  i b_{10} k_ik_jk_k A_{ij}^L U_k^L -\frac{1}{2} b_{10}^2 k_ik_j A_{ij}^L\xi_L  + \frac{1}{2} b_{20} \xi_L^2 
  -(b_{20}  + b_{10}^2) k_i  k_j  U_i^L U_j^L
   \nonumber\\
  &\qquad + 2i b_{10} b_{20} \xi_L k_i U_i^L 
  + i b_{10}^2 k_i U_i^{11} + i b_{20} k_i U_i^{20}- b_{10} k_ik_j A_{ij}^{10} \Bigg] \nonumber \\
  &\equiv P_A + P_W + P_{A^2} + P_{\xi_L} + P_{U} + P_{AU} + P_{A\xi_L} + P_{\xi^2_L} + P_{UU} + P_{U\xi_L} \label{PSPTintegrals} \\
   &\qquad + P_{U^{11}} + P_{U^{20}} + P_{A^{10}}, \nonumber
\end{align}
where we omit for the moment the curvature bias contributions.
We are searching for an expression involving only the $Q(k)$ and $R(k)$ functions of appendix \ref{app:kfunctions}.
Out of these integrals, the most cumbersome is $P_W$, that we compute  here: Using eq.~(\ref{wijk}) we have
$k_ik_jk_k W_{ijk} = k^3 (3 V(q) \mu + T(q)\mu^3)$ with  $\mu = \hat{k} \cdot \hat{q}$, hence
\begin{align}
 P_W(k) &= -\frac{i}{6} k_ik_jk_k \int d^3 q e^{ i \vk \cdot \vq} W_{ijk} = -\frac{i}{6} k^3 \int d^3q e^{i\vk \cdot \vq} (3 V(q) \mu + T(q)\mu^3) \nonumber\\
     &= -\frac{i}{6} k^3 2 \pi \int_0^\infty dq q^2 \int_{-1}^1 e^{ik q \mu}(3 V(q) \mu + T(q)\mu^3).
 \end{align}
Now, we use the identities $\int_{-1}^1 d\mu \mu e^{ix\mu} = 2i j_1(x)$ and $\int_{-1}^1 d\mu \mu^3 e^{ix\mu} = 
2i \left( \frac{3}{5}j_1(x) - \frac{2}{5}  j_3(x) \right)$, 
and from eqs.~(\ref{wijkV}, \ref{wijkT}) we have 
\begin{align}
 P_W &=   -\frac{2 \pi}{3} k^3 \frac{1}{\pi^2} \int^\infty_0 \frac{dp}{p}\int_0^\infty dq q^2 \left( 3 \tilde{V}(p) j_1(kq)j_1(pq)  
 - \frac{2}{5} \tilde{T}(p) j_3(kq) j_3(pq)  \right) \nonumber\\
     &=   -\frac{2}{3 \pi} k^3  \int^\infty_0 \frac{dp}{p}  3 \tilde{V}(p) \int_0^\infty dq q^2 j_1(kq)j_1(pq)
      +\frac{2}{3 \pi} k^3  \int^\infty_0 \frac{dp}{p}  \frac{2}{5} \tilde{T}(p) \int_0^\infty dq q^2 j_3(kq)j_3(pq) \nonumber\\
       &= -\tilde{V}(k) + \frac{2}{15}\tilde{T}(k).
 \end{align}
where we used the integral form of the Dirac delta function $  \dD(p-k) = \frac{2 p k}{\pi }\int_0^\infty dq q^2 j_\ell(kq)j_\ell(pq)$.
Using the definitions (\ref{Vtilde}, \ref{Ttilde}) we obtain
\begin{equation}
P_{W}(k) =   \frac{3}{7}Q_2(k) + \frac{6}{7}R_2(k).
\end{equation}

Analogous computations for the other integrals in eq.~(\ref{PSPTintegrals}) yield
\begin{align}
 P_A(k)           &= -\frac{1}{2}k_ik_j\int d^3 q e^{i \vk \cdot \vq } A_{ij}                
                   = P_L(k) + \frac{9}{98}Q_1(k) + \frac{10}{21}R_1(k), \\
 P_{A^2}(k)       &= \frac{1}{8}k_ik_jk_kk_l\int d^3 q e^{i \vk \cdot \vq } A^L_{ij} A^L_{kl}   
                   = \frac{1}{2}Q_3(k) - \sigma^2_L k^2P_L(k),\\
 P_{\xi_L}(k)     &= b_{10}^2 \int d^3 q e^{i \vk \cdot \vq } \xi_L     
                   =b_{10}^2 P_L(k),    \\
 P_{U}(k)         &= 2 b_{10} k_i \int d^3 q e^{i \vk \cdot \vq }  U_i      
                   = 2 b_{10} P_L(k) + \frac{10}{21} b_{10} R_1(k),\\
 P_{AU }(k)       &=-i b_{10} k_ik_jk_k \int d^3 q e^{i \vk \cdot \vq } A^L_{ij} U^L_k   
                   = 2b_{10} (Q_7(k)-k^2 \sigma^2_L  P_L(k)), \\
 P_{A\xi_L}(k)    &= -\frac{1}{2} b_{10}^2 k_ik_j \int d^3 q e^{i \vk \cdot \vq } A^L_{ij}\xi_L 
                   =b_{10}^2 (Q_{11}(k)- k^2 \sigma^2_L P_L(k)), \\
 P_{\xi_L^2}(k)   &=  \frac{1}{2} b_{20}\int d^3 q e^{i \vk \cdot \vq } \xi_L^2 
                   = \frac{1}{2}  b_{20}^2  Q_{13}(k) \label{Q13der}, \\
 P_{UU}(k)        &= -(b_{20} + b_{10}^2) k_i  k_j \int d^3 q e^{i \vk \cdot \vq } U^L_iU^L_j
                   = ( b_{20}+ b_{10} ^2)  Q_{9}(k), \\  
 P_{U\xi_L}(k)    &=  2i b_{10} b_{20}k_i \int d^3 q e^{i \vk \cdot \vq } \xi_L  U^L_i 
                   = 2 b_{20}b_{10}  Q_{12}(k)  \\ 
 P_{U^{11}}(k)    &=  i b_{10}^2 k_i\int d^3 q e^{i \vk \cdot \vq } U_i^{11}
                   = \frac{6}{7} b_{10}^2 R_{1+2}(k), \\
 P_{U^{20}}(k)    &= i b_{20}k_i \int d^3 q e^{i \vk \cdot \vq } U_i^{20}
                   = \frac{3}{7} b_{20}Q_8(k),     \\
 P_{A^{10}}(k)    &= - b_{10} k_ik_j \int d^3 q e^{i \vk \cdot \vq } A_{ij}^{10}
                   =\frac{6}{7}b_{10}(R_{1+2}(k) + R_2(k) + Q_5(k)). 
\end{align}
Considering the terms coming from trivial integrals of $\nabla^2 \xi_L$ and $\nabla^4 \xi_L$, 
giving $-k^2 P_L$ and $k^4 P_L$ respectively, and rearranging terms,  we arrive to eq.~(\ref{XSPTPS}).

\end{section}

\acknowledgments

A.A and J.L.C-C  acknowledge partial support by Conacyt Fronteras Project 281. 
A.A, J.L.C-C and M.A.R.M acknowledge partial support by Conacyt project 283151.



 \bibliographystyle{JHEP}  
 \bibliography{biasMGbib.bib}  
 
\end{document}